\documentclass[11pt]{article}
\pdfoutput=1
\usepackage{graphicx}
\usepackage{amsmath}
\usepackage{amssymb}
\usepackage{epsfig}
\usepackage{subeqnarray}
\usepackage{color}

\title{Efficient simulation of multidimensional phonon transport using  energy-based variance-reduced Monte Carlo formulations}
\author{Jean-Philippe M. P\'eraud and Nicolas G. Hadjiconstantinou\\Department of Mechanical Engineering, Massachusetts Institute of Technology\\ Cambridge, MA  02139, USA}

\begin{document}

\maketitle

\begin{abstract}
We present a new Monte Carlo method for obtaining solutions of the Boltzmann equation for describing phonon  transport in micro and nanoscale devices. 
The proposed method can resolve arbitrarily small signals (e.g. temperature differences) at small constant cost and thus represents a considerable improvement 
compared to traditional Monte Carlo methods whose cost increases quadratically with decreasing signal. This is achieved via a control-variate 
variance reduction formulation in which the stochastic particle description only solves for the deviation from a nearby equilibrium, while the latter 
is described analytically. We also show that simulating an energy-based Boltzmann equation results in an algorithm that lends itself naturally to exact energy 
conservation thereby considerably improving the simulation fidelity. Simulations using the proposed method are used to investigate the 
effect of porosity on the 
effective thermal conductivity of silicon. We also present simulations of a recently developed thermal conductivity spectroscopy process. The latter 
simulations demonstrate how the computational gains introduced by the proposed method enable the simulation of otherwise intractable multiscale phenomena.  

\end{abstract}
\section{Introduction}\label{introsec}

Over the past two decades, the dramatic advances associated with MEMS (Micro Electro Mechanical Systems) 
and NEMS (Nano Electro Mechanical Systems) have attracted considerable attention on microscale and nanoscale heat transfer considerations \cite{chenbook}. 
Applications range from thermal management of electronic devices \cite{Majumdar93} to the development of thermoelectric materials with higher figure of merit \cite{Chen03}. 
The thermoelectric figure of merit is proportional to the electrical conductivity and inversely proportional to thermal 
conductivity and can thus be improved by reducing the latter and/or increasing the former. One of the most promising approaches towards reducing the 
thermal conductivity of thermoelectric materials is the introduction of nanostructures that interact with the ballistic 
motion of phonons at small scales thus influencing heat transport \cite{Jeng08}. 
Such approach requires a reliable description of phonon transport at the nanoscale and cannot rely on  
Fourier's law, which is valid for diffuse transport. On the other hand, first principles calculations 
(e.g. molecular dynamics approaches, classical or quantum mechanical) are too expensive for treating phonon transport at the device (e.g. micrometer) scale. 
At these scales, a kinetic description based on the Boltzmann Transport Equation (BTE) offers a reasonable balance between fidelity and model complexity and is able to 
accurately describe the transition from diffusive to ballistic transport as characteristic system lengthscales approach and ultimately  become smaller than the phonon mean free path. 

Solving the BTE is a challenging task, especially in complex geometries. The high dimensionality of the 
distribution function coupled with the ability of particle methods to naturally simulate advection processes without stability problems \cite{Baker05} 
make particle Monte Carlo methods particularly appealing. Following the development of the Direct Monte Carlo Method by Bird \cite{bird} for treating dilute 
gases, Monte Carlo methods for phonon transport were first 
introduced by Peterson \cite{Peterson94} and subsequently improved by Mazumder and Majumdar \cite{Mazumder01}. Over the past decade, further important 
refinements have been introduced: Lacroix \textit{et al.} introduced a method to treat frequency dependent mean free paths \cite{Lacroix05}; Jeng \textit{et al.} introduced 
a method for efficiently treating transmission and reflection of phonons at material interfaces and used this method to model the thermal conductivity of 
nanoparticle composites \cite{Jeng08}. Hao \textit{et al.} developed \cite{Hao09} a formulation for periodic boundary conditions 
in order to study the thermal conductivity of periodic nanoporous materials while only simulating one unit cell (period).

The work presented here introduces a number of improvements which enable efficient and accurate simulation of the most challenging phonon transport 
problems, namely three-dimensional and transient. Accuracy is improved compared to previous approaches by 
introducing an energy-based formulation, which simulates energy packets rather than phonons; this formulation makes energy conservation particularly easy to implement 
rigorously, in contrast to previous approaches which were ad-hoc and in many cases ineffective. 
We also introduce a variance-reduced formulation for {\it substantially} reducing the statistical uncertainty associated with sampling solution 
(temperature and heat flux) fields. This formulation is based on the concept of control variates, first introduced  
in the context of Monte Carlo solutions of the Boltzmann equation for dilute gases \cite{Baker05}; it is based on  the fact that signal 
strength is intimately linked to deviation from equilibrium, or in other words, that the large computational cost associated with small signals 
is due to the fact that in these problems the deviation from equilibrium is small. 
This observation can be exploited by utilizing the nearby equilibrium state as a ``control'' and using the Monte Carlo method to calculate the contribution 
of non-equilibrium therefrom. Because the deviation from equilibrium is small, only a small quantity is evaluated stochastically 
(the fields associated with the equilibrium component are known analytically) resulting in small statistical uncertainty; 
moreover, the latter decreases as the deviation from equilibrium decreases, thus enabling the simulation of arbitrarily 
small deviations from equilibrium. 

In the technique presented here, we use particles to simulate the deviation from equilibrium, which is thus referred to as a deviational method; 
the origin of this methodology can be  found in the Low Variance Deviational Simulation Monte Carlo (LVDSMC) method \cite{homollea07,homolleb07,radtke09,radtke11} recently 
developed for dilute gases. The theoretical basis underlying this method as well as the modifications required for use in phonon transport simulations 
are described in section \ref{subsec:deviational_formulation}. The resulting algorithm is described in section \ref{vrsec} and  validated in section \ref{sec:validation}. 

The proposed algorithm is used to obtain solutions to two problems of practical interest. The first application studies the thermal 
conductivity of porous silicon containing voids with different degrees of alignment and is intended to showcase how ballistic effects influence the ``effective'' 
thermal conductivity. The second application is related to the recently developed experimental method of ``thermal conductivity spectroscopy'' \cite{minnich} based on the pump-probe technique known as transient thermoreflectance, which uses 
the response of a material to laser irradiation to infer information about physical properties of interest \cite{Koh07} (e.g. mean free paths of the dominant heat carriers).

\section{Theoretical basis}\label{theorysec}
\subsection{Summary of traditional Monte Carlo simulation methods}
\label{subsec:usual_MC}
We consider the Boltzmann Transport Equation in the frequency-dependent relaxation-time approximation  
\begin{equation}
\frac{\partial f}{\partial t}+\mathbf{V}_{g}(\omega,p)\nabla f =-\frac{f-f^{loc}}{\tau(\omega,p,T)}
\label{eq:BTE}
\end{equation}
where, $f=f(t,\mathbf{r},\mathbf{k},p)$ is the phonon distribution in the phase space, 
$\omega=\omega(\mathbf{k},p)$ the phonon radial frequency, $p$ the phonon polarization and $T$ the temperature; 
similarly to the nomenclature adopted in \cite{chenbook}, $f$ is defined in reference to the occupation number. 
For example, if the system is perfectly thermalized at temperature $T_{eq}$, $f$ is a Bose-Einstein distribution
\begin{equation}
f^{eq}_T=\frac{1}{\exp \left( \frac{\hbar \omega(\mathbf{k},p)}{k_{b} T} \right) -1}
\end{equation}
where $k_b$ is Boltzmann's constant.
Also, $f^{loc}$ is an equilibrium (Bose-Einstein) distribution corresponding to the {\it local pseudo-temperature} 
defined more precisely in section \ref{ti}. 

In this work we consider Longitudinal Acoustic (LA), Transverse Acoustic (TA), Longitudinal Optical (LO), Transverse Optical (TO) polarizations; 
acoustic phonons are 
known to be the most important contributors to lattice thermal conductivity \cite{Klemens58,Mittal10}. 
The phonon radial frequency is given by the dispersion relation $\omega=\omega(\mathbf{k}, p)$.
Phonons travel at the group velocity $\mathbf{V}_{g}=\nabla_{k} \omega$. 

In the following, we always consider the ideal case where the dispersion relation is isotropic. For convenience, the radial frequency $\omega$ 
and two polar angles $\theta$ and $\phi$ are usually preferred as primary parameters compared to the wave vector.
Equation (\ref{eq:BTE}) is simulated using computational particles that represent phonon bundles, namely collections of phonons with similar characteristics 
(position vector $\mathbf{x}$, the wave vector $\mathbf{k}$, and the polarization/propagation-mode $p$), using the approximation

\begin{equation}
\frac{1}{8 \pi^3}f(t,\mathbf{x},\mathbf{k},p) \approx N_{eff}\sum_{i} \delta^3(\mathbf{x}-\mathbf{x}_{i})\delta^3(\mathbf{k}-\mathbf{k}_i)\delta_{p,p_i}
\label{eq:phonon_bundles}
\end{equation}
where $\mathbf{x}_{i}$, $\mathbf{k}_{i}$ and $p_{i}$ respectively represent the position, the wave vector and the 
polarization of particle $i$ and $N_{eff}$ is the number of phonons in each phonon bundle. The factor $1/8\pi^3$ is 
necessary for converting the quantity representing the occupation number, $f$, into a quantity representing the phonon 
density in phase space. Written in polar coordinates, and using the frequency instead of the wave number, this expression becomes
\begin{equation}
\frac{D(\omega,p)}{4\pi}f(t,\mathbf{x},\omega,\theta,\phi,p) \sin (\theta) \approx N_{eff}\sum_{i} \delta^3(\mathbf{x}-\mathbf{x}_{i})\delta(\omega-\omega_{i})\delta(\theta-\theta_i)\delta(\phi-\phi_i)\delta_{p,p_i}
\label{eq:phonon_bundles2}
\end{equation}
where $\omega_{i}$, $\theta_i$, and $\phi_i$ respectively represent the radial frequency, the polar angle and the azimuthal angle of particle $i$. 
The density of states, $D(\omega,p)$, is made necessary by the use of $\omega$ as a primary parameter and is given by
\begin{equation}
D(\omega,p)=\frac{k(\omega,p)^2}{2\pi^2 V_{g}(\omega,p)}
\end{equation}

\subsubsection{Initialization}
Systems are typically initialized in an equilibrium state at temperature $T$; the number of 
phonons in a given volume V is calculated using the Bose-Einstein statistics 
\begin{equation}
N=V\int_{\omega=0}^{\omega_{max}}\sum_{p} D(\omega,p) f^{eq}_{T}(\omega)d\omega
\end{equation}
where:
\begin{itemize}
\item $\omega_{max}$ is the maximum (cut-off) frequency
\item $f^{eq}_{T}$ is the occupation number at equilibrium at temperature $T$
\end{itemize}
The number of computational particles (each representing a phonon bundle) is given by $N/N_{eff}$. 
The value of $N_{eff}$ is determined by balancing computational cost (including storage) with the need for a sufficiently large number of particles 
for statistically meaningful results.

\subsubsection{Time integration}
\label{ti}
Once the system is initialized, the simulation proceeds by applying a splitting algorithm with timestep $\Delta t$. Integration for one timestep comprises of three substeps:
\begin{itemize}
\item The advection substep during which bundle $i$ moves by $\mathbf{V}_{g,i} \Delta t$.
\item The sampling substep during which the temperature $(T)$ and pseudo-temperature $(T_{loc})$ are locally measured. 
They are calculated by inverting the local energy $(E)$ and pseudo-energy $(\tilde{E})$ \cite{Hao09} relations
\begin{equation}
E=N_{eff}\sum_{i} \hbar \omega_{i}= V \int_{\omega=0}^{\omega_{max}} \sum_{p}  \frac{D(\omega,p)\hbar \omega}{\exp\left( \frac{\hbar \omega}{k_{b} T} \right) -1} d\omega
\end{equation}
and
\begin{equation}
\tilde{E}=N_{eff}\sum_{i} \frac{\hbar \omega_{i}}{\tau(\omega_{i},p_i,T)}=V \int_{\omega=0}^{\omega_{max}} \sum_{p}  \frac{D(\omega,p)\hbar \omega}{\tau(\omega,p,T)}\frac{1}{\exp\left( \frac{\hbar \omega}{k_{b} T_{loc}} \right) -1} d\omega
\end{equation}
respectively.
\item The scattering substep, during which each phonon $i$ is scattered according to its scattering probability given by
\begin{equation}
P_i=1-\exp\left( -\frac{\Delta t}{\tau(\omega_i,p_i,T)} \right)
\end{equation}
Scattering proceeds by drawing new frequencies, polarizations and traveling directions. Because of the frequency dependent relaxation times, frequencies must be drawn 
from the distribution $D(\omega,p)f^{loc}/\tau(\omega,p,T)$.
Since scattering events conserve energy, the latter must be conserved during this substep. However, because the 
frequencies of the scattered phonons are drawn randomly, 
conservation of energy is enforced by adding or deleting particles until a 
target energy is approximately reached \cite{Mazumder01,Lacroix05}. In addition to being approximate, this method does not always ensure that energy is conserved, resulting 
in random walks in the energy of the simulated system, which  in some cases leads to deterministic error. In the next section, we present a convenient way for rigorously conserving energy.
\end{itemize}

\subsection{Energy based formulation} 

While most computational techniques developed so far only conserve energy in an approximate manner \cite{Mazumder01, Lacroix05}, here we 
show that an energy-based formulation provides a convenient and rigorous way to conserve energy in the relaxation time approximation. 

Adopting a similar approach as in \cite{Majumdar93} to derive the {\it Equation of Phonon Radiative Transfer}, we  multiply (\ref{eq:BTE}) by $\hbar \omega$ to obtain 
\begin{equation}
\frac{\partial e}{\partial t} + \mathbf{V}_{g} \nabla e = \frac{e^{loc}-e}{\tau}
\label{eTE}
\end{equation}
which we will refer to as the energy-based BTE. Here, $e=\hbar \omega f$ and  $e^{loc}=\hbar \omega f^{loc}$. 
Equation (\ref{eTE}) can be simulated by writing
\begin{equation}
e \approx 8\pi^3 \mathcal{E}_{eff}\sum_{i} \delta^3(\mathbf{x}-\mathbf{x}_{i})\delta^3(\mathbf{k}-\mathbf{k}_{i})\delta_{p,p_i}
\label{eq:energetic_bundles}
\end{equation}
where $\mathcal{E}_{eff}$ is defined as the effective energy carried by each computational particle. Statement (\ref{eq:energetic_bundles}) defines computational 
particles that all represent the same amount of energy. From the point of view of phonons, comparing (\ref{eq:phonon_bundles}) and (\ref{eq:energetic_bundles}) shows 
that the effective number of phonons represented by the newly defined particles is variable and is linked to the effective energy by the relation 
$\mathcal{E}_{eff}=N_{eff} \hbar \omega$.
By analogy with the description of section \ref{subsec:usual_MC}, computational particles defined by (\ref{eq:energetic_bundles}) obey the same computational 
rules as in the previous Monte Carlo approaches. Modifications appear at three levels:
\begin{itemize}
\item When drawing particle frequencies during initialization, emission from boundaries or scattering, the distribution functions that we use must account 
for the factor $\hbar \omega$. For example, when initializing an equilibrium population of particles at a temperature $T$ , one has to draw the frequencies from the distribution
\begin{equation}
\frac{\hbar \omega \sum_p D(\omega,p)}{\exp \left( \frac{\hbar \omega}{k_{b}T}\right) -1}
\end{equation}
\item Calculating the energy in a cell is straightforward and simply consists in counting the number of computational particles. 
The energy associated with $N$ particles is given by $\mathcal{E}_{eff}N$.
\item Since the energy in a cell is proportional to the number of particles, there is no need for an addition/deletion process: 
{\it energy is strictly and automatically conserved by simply conserving the number of particles}.
\end{itemize}

\subsection{Deviational formulation}
\label{subsec:deviational_formulation}
\begin{figure}[htbp]
	\centering
	\includegraphics[width=.5\textwidth]{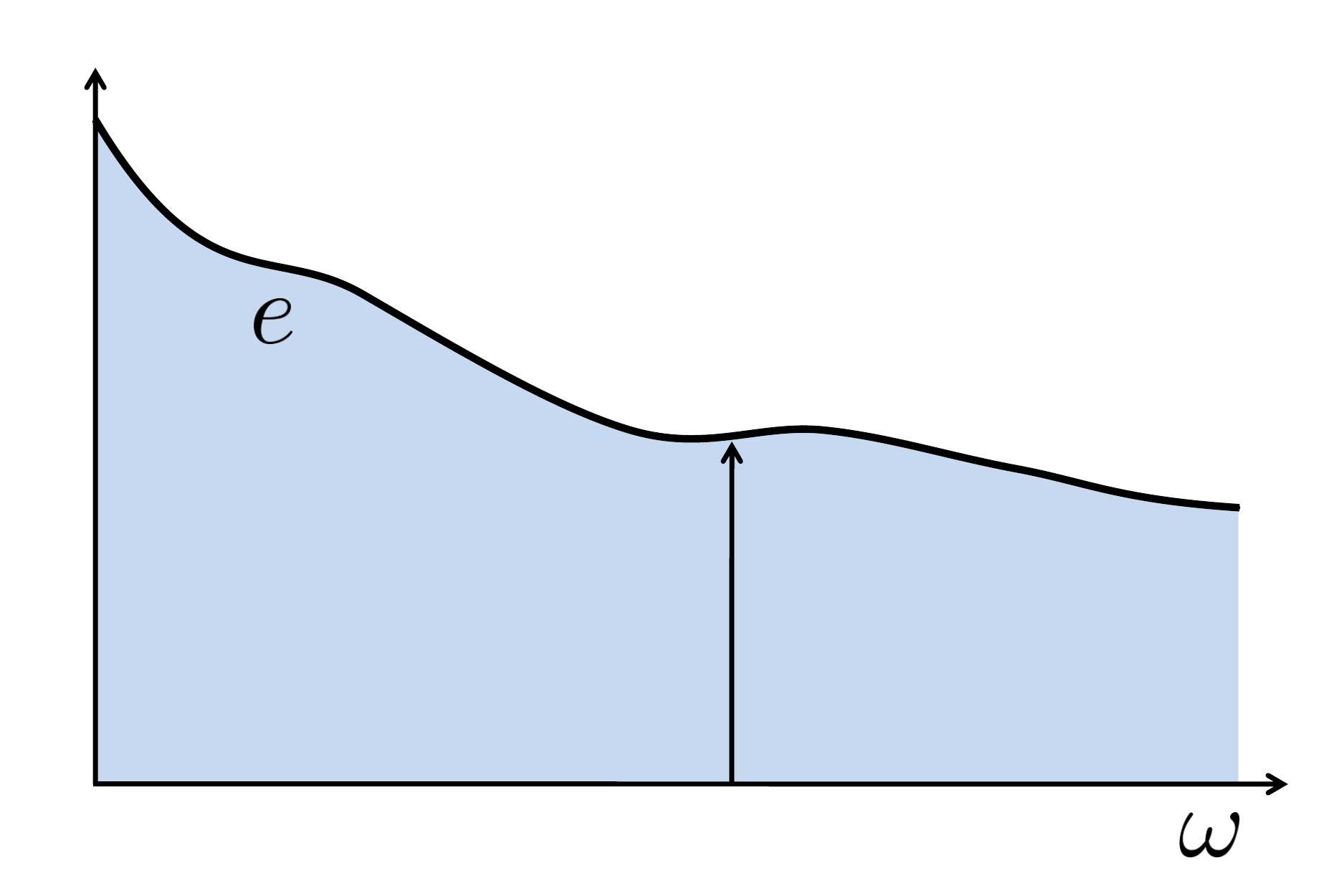}
	\caption{In standard particle methods, the moments of the distribution are stochastically integrated.}
	\label{fig:control_V1}
\end{figure}

\begin{figure}[htbp]
	\centering
	\includegraphics[width=.5\textwidth]{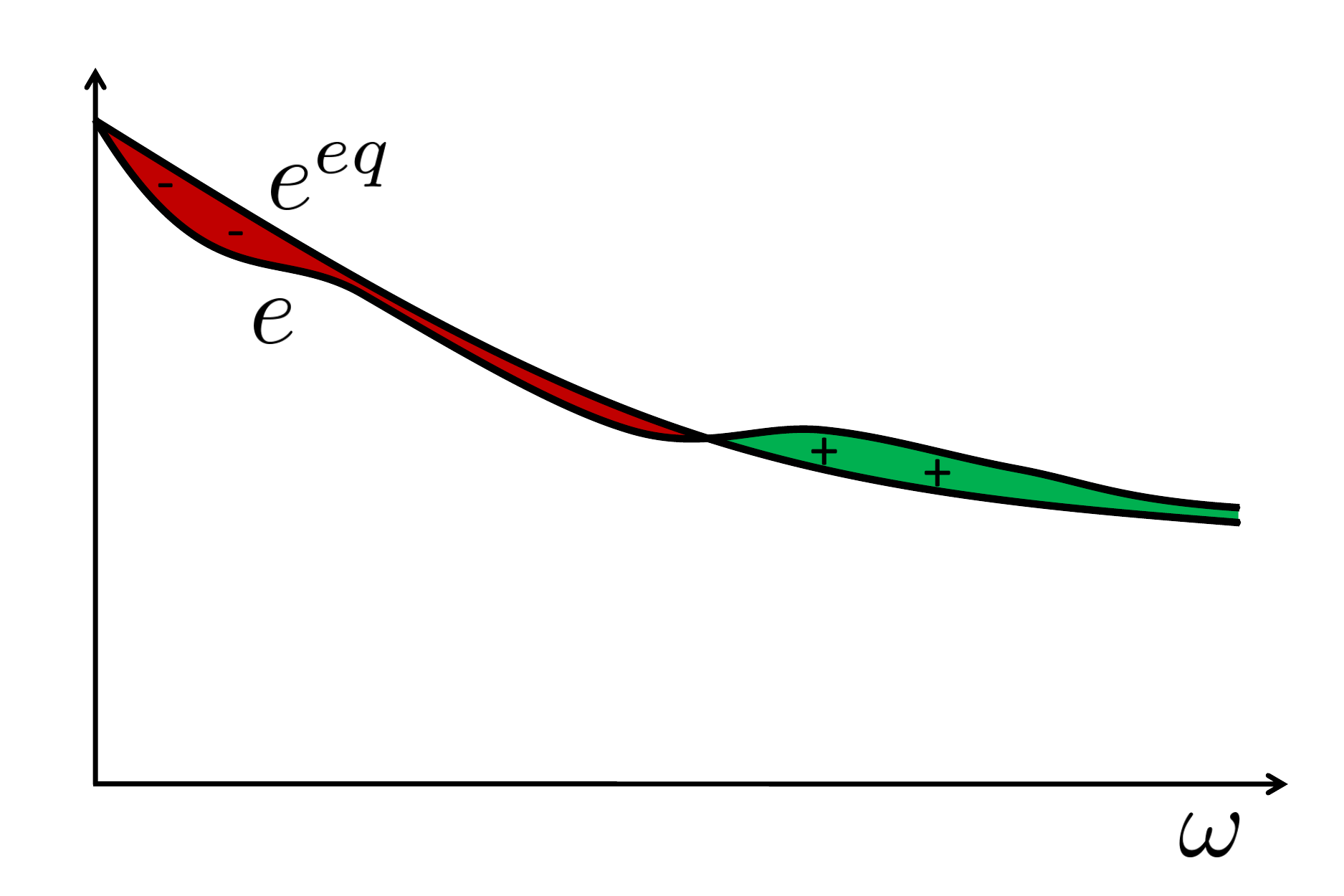}
	\caption{In a control-variate formulation, the stochastic part is reduced to the calculation 
of the deviation from a known state, which is much smaller.}
	\label{fig:control_V2}
\end{figure}

In this section we introduce an additional modification which dramatically decreases the statistical uncertainty associated with Monte Carlo simulations of (\ref{eTE}).
Our approach belongs to a more general class of control-variate variance reduction methods for solving kinetic equations \cite{Baker05,homollea07,radtke} 
in which the moments $<R>$ of a given distribution $f$ are computed by writing
\begin{equation}
\int R f d\mathbf{x}d\mathbf{c} = \int R (f-f^{eq}) d\mathbf{x}d\mathbf{c}+\int R f^{eq} d\mathbf{x}d\mathbf{c}
\end{equation}
where the first term of the right hand side is computed stochastically and the second term is computed deterministically. If $f^{eq}\approx f$, the variance 
reduction is large because only a small term is determined stochastically (see Figures \ref{fig:control_V1} and \ref{fig:control_V2}).

In the present context, this methodology provides significant computational savings when an equilibrium (constant temperature) state exists nearby, which is 
precisely the regime in which statistical noise becomes problematic (low signals). The degree of variance reduction achieved by this method is 
quantified in section \ref{sec:efficiency}.

Let
\begin{equation}
e^{eq}_{T_{eq}}(\omega)=\frac{\hbar \omega}{\exp \left(\frac{\hbar \omega}{k_{b} T_{eq}}\right)-1}
\end{equation}
where $T_{eq}\neq T_{eq}(\mathbf{x},t)$. Then, it is straightforward to show that $e^d=e-e^{eq}_{T_{eq}}$ is governed by 
\begin{equation}
\frac{\partial e^d}{\partial t} + \mathbf{V}_{g} \nabla e^d = \frac{(e^{loc}-e^{eq}_{T_{eq}})-e^d}{\tau}
\label{eq:BTE_VR}
\end{equation}

Therefore, by analogy to the standard particle methods for solving the Boltzmann equation, we define computational particles by:
\begin{equation}
e^d=e-e^{eq}_{T_{eq}} \approx 8\pi^3 \mathcal{E}^{d}_{eff} \sum_i s(i) \delta^3(\mathbf{x}-\mathbf{x}_{i})  \delta^3(\mathbf{k}-\mathbf{k}_{i}) \delta_{p,p_i}, \ s(i)=\pm 1
\label{eq:deviational_particles}
\end{equation}
We will refer to these newly defined computational particles as deviational particles. Clearly, deviational particles may be negative since 
$e-e^{eq}_{T_{eq}}$ can be a negative quantity. This is accounted for in the sign term in equation (\ref{eq:deviational_particles}). In what follows, 
we derive evolution rules for deviational particles based on (\ref{eq:BTE_VR}). 

\section{Algorithm}\label{vrsec}
The variance-reduced algorithm is very similar to its non-variance reduced counterpart and comprises an initialization step followed by a splitting 
algorithm for time integration. The main change lays in the distributions from which deviational particles are sampled.

\subsection{Initialization}
\label{subsec:init}
The algorithm proceeds by choosing the equilibrium state at temperature $T_{eq}$ from which deviations will be simulated. Although this choice can be quite critical 
in the efficiency of the method (the smaller the deviation from the chosen equilibrium state, the smaller the number of deviational particles required for a 
given statistical uncertainty, or for a fixed number of deviational particles, the larger the variance reduction), it is usually a natural and intuitive choice.

In some cases, the equilibrium state is the same as the initial state. In such a situation, the simulation starts with no particles. 
Nevertheless, one still has to choose the deviational effective energy $\mathcal{E}^{d}_{eff}$ for subsequent use. In the various examples discussed below, this parameter was chosen as follows:
based on a guess of the upper bound on the deviation of temperature at steady state, the deviational energy of the system can be estimated using
\begin{equation}
\Delta E=\int_{\omega=0}^{\omega_{max}} \sum_{p} \hbar \omega D(\omega,p) \left| \frac{1}{\exp\left( \frac{\hbar \omega}{k_{b} T}\right) -1}-\frac{1}{\exp\left( \frac{\hbar \omega}{k_{b} T_{eq}}\right) -1} \right| d\omega
\end{equation}
This estimate of the deviational energy allows $\mathcal{E}^{d}_{eff}$ to be (approximately) determined based on the desired  number of computational particles.

If the initial state $f^0$ is different from the equilibrium distribution, particles need to be initialized in the computational domain. Their frequencies and polarizations 
are drawn from the distribution
\begin{equation}
D(\omega,p)e^{d}(\omega) =\hbar \omega D(\omega,p) \left[ f^0-\frac{1}{\exp\left( \frac{\hbar \omega}{k_{b} T_{eq}}\right) -1} \right]
\label{energy_deviational_distribution_0}
\end{equation}
Typically, $f^0$ is an equilibrium distribution at some temperature $T$, whereby the above expression reduces to
\begin{equation}
D(\omega,p)e^{d}(\omega) =\hbar \omega D(\omega,p) \left[ \frac{1}{\exp\left( \frac{\hbar \omega}{k_{b} T}\right) -1}-\frac{1}{\exp\left( \frac{\hbar \omega}{k_{b} T_{eq}}\right) -1} \right]
\label{energy_deviational_distribution}
\end{equation}
This function is positive if $T>T_{eq}$ and negative if $T<T_{eq}$. As a result, in the latter case, particles are assigned a negative sign. Drawing the frequencies is performed as in \cite{Mazumder01}, namely by subdividing the frequency range in bins (generally, about 1000 bins are considered enough), defining a discretized and normalized cumulative distribution from (\ref{energy_deviational_distribution}), uniformly drawing a random number between 0 and 1 and finding the bins it corresponds to in order to match the normalized cumulative distribution.

\subsection{Advection}
Since the left hand side of (\ref{eq:BTE_VR}) is analogous to that of (\ref{eq:BTE}), the advection substep is unchanged. In other words, 
during the time step $\Delta t$, particles of group velocity $\mathbf{V}_{g}(\omega,p)$ are simply advected by $\mathbf{V}_{g}(\omega,p) \Delta t$.

\subsection{Sampling substep}
Sampling the local temperature and pseudo-temperature requires a few changes from the non-variance reduced method, namely
\begin{itemize}
\item Let $C_{j}$ be the set of indexes corresponding to the particles inside cell $j$ of volume $V_j$ at time $t$. Since each particle represents the same amount of energy, 
the deviational energy is given by
\begin{equation}
\Delta E_{j} =\mathcal{E}^{d}_{eff} \sum_{i \in C_j} s(i)=\mathcal{E}^{d}_{eff} (\mathcal{N}^{+}_j-\mathcal{N}^{-}_j)
\end{equation}
where $\mathcal{N}^{+}_j$ and $\mathcal{N}^{-}_j$ are respectively the number of positive and negative particles inside the cell $j$.
\item The corresponding temperature $T_j$ is then calculated by numerically inverting the expression
\begin{equation}
\frac{\Delta E_{j}}{V_{j}} = \int_{\omega=0}^{\omega_{max}} \sum_{p} D(\omega,p)\hbar \omega \left[ \frac{1}{\exp\left( \frac{\hbar \omega}{k_{b} T_{j}} \right) -1} - \frac{1}{\exp \left(  \frac{\hbar \omega}{k_{b} T_{eq}} \right) -1} \right] d\omega
\end{equation}
\item Similarly, once $T_{j}$ is known, the deviational pseudo-energy is computed using 
\begin{equation}
\Delta \tilde{E}_j = \mathcal{E}^{d}_{eff} \sum_{i \in C_j} \frac{s(i)}{\tau(\omega_i,p_i,T_{j})}
\end{equation}
\item The corresponding pseudo-temperature $[T_{loc}]_j$ is calculated by numerically inverting
\begin{equation}
\frac{\Delta \tilde{E}_{j}}{V_{j}} = \int_{\omega=0}^{\omega_{max}} \sum_{p} \frac{D(\omega,p)\hbar \omega}{\tau(\omega,p,T_{j})} \left[ \frac{1}{\exp\left( \frac{\hbar \omega}{k_{b} [T_{loc}]_j} \right) -1} - \frac{1}{\exp \left(  \frac{\hbar \omega}{k_{b} T_{eq}} \right) -1} \right] d\omega
\end{equation}
\end{itemize}

\subsection{Scattering step}
\label{subsec:scattering_step}
During the scattering step we integrate 
\begin{equation}
\frac{de^{d}}{dt}=\frac{(e^{loc}-e^{eq}_{T_{eq}})-e^d}{\tau(\omega,p,T_{j})}
\end{equation}
for a timestep $\Delta t$, where
\begin{equation}
e^{loc}-e^{eq}_{T_{eq}}=\hbar \omega \left[\frac{1}{\exp \left( \frac{\hbar \omega}{k_{b} [T_{loc}]_j}\right)-1}-\frac{1}{\exp \left( \frac{\hbar \omega}{k_{b} T_{eq}}\right)-1}\right]
\end{equation}
We select the particles to be scattered according to the scattering probability (specific to each particle's frequency and polarization, and depending on the local 
temperature)
\begin{equation}
P(\omega_i,p_i,T_j)=1-\exp \left(-\frac{\Delta t}{\tau(\omega_i,p_i,T_j)}\right)
\end{equation}
The pool of selected particles represent a certain amount of deviational energy 
$\mathcal{E}^d_{eff}(\mathcal{N}^{+}_{s,j}-\mathcal{N}^{-}_{s,j})$, where $\mathcal{N}^{+}_{s,j}$ 
and $\mathcal{N}^{-}_{s,j}$ refer respectively to the number of positive and negative selected (i.e. scattered) 
particles in cell $j$. 
This pool of selected particles must be replaced by particles with properties drawn from the distribution
\begin{equation}
\frac{D(\omega,p)(e^{loc}-e^{eq}_{T_{eq}})}{\tau(\omega,p,T_j)}=\frac{D(\omega,p)\hbar \omega}{\tau(\omega,p,T_j)} \left( \frac{1}{\exp \left( \frac{\hbar \omega}{k_b [T_{loc}]_j} \right)-1}-\frac{1}{\exp \left( \frac{\hbar \omega}{k_b {T}_{eq}} \right)-1} \right)
\label{eq:scattering}
\end{equation}
which is either positive for all frequencies and polarizations or negative for all frequencies and polarizations. 
In other words, scattered particles must be replaced by particles which all have the same sign as $e^{loc}-e^{eq}_{T_{eq}}$ and which respect the energy conservation requirement. 
Therefore, out of the $\mathcal{N}^{+}_{s,j}+\mathcal{N}^{-}_{s,j}$ selected particles, we redraw properties for 
$\left| \mathcal{N}^{+}_{s,j}-\mathcal{N}^{-}_{s,j} \right|$ of them according to the distribution (\ref{eq:scattering}) and delete the other selected particles. The $\left| \mathcal{N}^{+}_{s,j}-\mathcal{N}^{-}_{s,j} \right|$ 
particles to be kept are chosen randomly inside the cell $j$ and are given the sign of $e^{loc}-e^{eq}_{T_{eq}}$.

This process tends to reduce the number of particles in the system and counteracts sources of particle creation within the algorithm (e.g. see boundary 
conditions discussed in the next section). A bounded number of particles is essential to the method stability and 
the reduction process just described is a major contributor to the latter \cite{homollea07,homolleb07}. 
Hence, in a typical problem starting from an equilibrium state that is also chosen as the control, the number of particles will first increase from zero 
and, at steady state, reach a constant value that can be estimated by appropriately choosing $\mathcal{E}^{d}_{eff}$ as described in section \ref{subsec:init}. The constant value will usually be higher (but of the same order) than the estimated value: indeed, the rate of elimination of pairs of particles of opposite signs depends on the number of particles per cell and therefore on the spatial discretization chosen (the finer the discretization, the smaller the number of particles per cell and therefore the smaller the rate of elimination).

\subsection{Boundary conditions}
\label{subsec:BC}
In phonon transport problems, various types of boundary conditions appear. Isothermal boundary conditions, similar by nature to a black body, 
have been used in several 
studies \cite{Mazumder01,Lacroix05}. Adiabatic boundaries also naturally appear \cite{Mazumder01,Lacroix06}. Recently, a class of periodic boundary conditions has also been  
introduced \cite{Hao09}. The deviational formulation adapts remarkably well to these different classes of boundary conditions.

\subsubsection{Adiabatic boundaries}
Adiabatic boundaries reflect all incident phonons. This reflection process can be divided into two main categories: diffuse reflection and specular reflection. In both cases, it is assumed that the polarization and frequency remains the same when a phonon is reflected. The only modified parameter during the process is the traveling direction.
\begin{itemize}
\item[i] \textit{Specular reflection} on a boundary $\partial n$ of normal vector $\mathbf{n}$ can be expressed, in terms of energy distribution, by
\begin{equation}
e(\mathbf{x},\mathbf{k})=e(\mathbf{x},\mathbf{k}')
\end{equation}
where $\mathbf{k}'=\mathbf{k}-2(\mathbf{k}\cdot \mathbf{n})\mathbf{n}$ and $\mathbf{x} \in \partial n$.
Since the equilibrium distribution $e^{eq}_{T_{eq}}$ is isotropic, then substracting it from both sides simply yields
\begin{equation}
e^d(\mathbf{x},\mathbf{k})=e^d(\mathbf{x},\mathbf{k}')
\end{equation}
In other words,  deviational particles are specularly reflected
\item[ii] \textit{Diffuse reflection} amounts to randomizing the traveling direction of a phonon incident on the boundary, in order for the population of 
phonons leaving the boundary to be isotropic. Since an equilibrium distribution is already isotropic, incident deviational particles are treated 
identically to real phonons.

\end{itemize}

\subsubsection{Isothermal boundaries}
\label{iso}
In the case of an isothermal boundary at temperature $T_b$, incident phonons are absorbed, while the boundary itself, 
at temperature $T_{b}$, 
emits new phonons from the equilibrium distribution corresponding to $T_{b}$. The emitted heat flux per unit radial frequency is expressed by
\begin{equation}
q''_{\omega,b}= \frac{1}{4}\sum_p \frac{D(\omega,p)V_{g}(\omega,p) \hbar \omega}{\exp \left( \frac{\hbar \omega}{k_b T_b} \right) -1}
\end{equation}

Substracting the heat flux per unit radial frequency corresponding to a boundary at equilibrium temperature, we obtain 
\begin{equation}
q''_{\omega,b}= \frac{1}{4}\sum_p D(\omega,p)V_{g}(\omega,p) \hbar \omega \left( \frac{1}{\exp \left( \frac{\hbar \omega}{k_b T_b} \right) -1}-\frac{1}{\exp \left( \frac{\hbar \omega}{k_b T_{eq}} \right) -1} \right)
\end{equation}
which gives the frequency distribution of emitted particles. Traveling directions must be chosen accordingly, as explained for example in \cite{Mazumder01}.

\subsubsection{Periodic unit cell boundary conditions}
\label{subsubsec:periodic_boundaries}
Heat transfer in periodic nanostructures is a subject of considerable interest in the context of many applications. 
Such nanostructures are  considered in Hao \textit{et al.} \cite{Hao09}, in Huang \textit{et al.} \cite{Huang09} and in Jeng \textit{et al.} \cite{Jeng08}. 
Hao \textit{et al.} developed periodic boundary conditions that allow efficient simulation of such structures by considering only one unit cell (period). 
In this section we review the work of Hao \textit{et al.} \cite{Hao09} and explain how the deviational particle formulation presented here lends itself naturally to this type of 
boundary condition. Simulations using these boundary conditions are presented in section \ref{subsec:staggered-aligned}.

\begin{figure}[htbp]
	\centering
	\includegraphics[width=.6\textwidth]{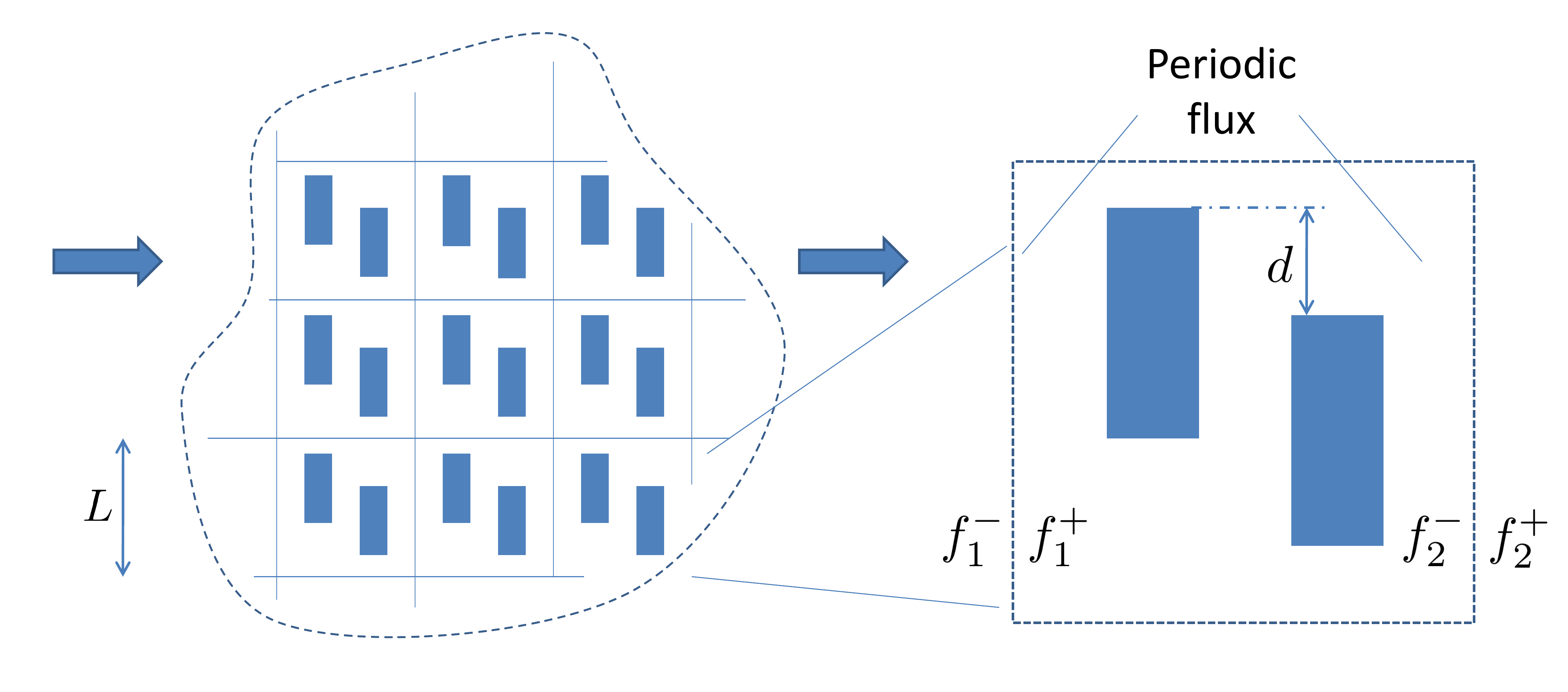}
	\caption{Example of a periodic nanostructure. Each periodic cell comprises two rectangular voids with diffusely reflecting walls. This nanostructure, and in particular the influence of the parameter $d$, is studied further in section \ref{subsec:staggered-aligned}}
	\label{fig:periodic_struct_bc2}
\end{figure}

We consider a 2D periodic structure depicted in Figure \ref{fig:periodic_struct_bc2} in which square unit cells containing two rectangular voids are organised 
in a square lattice. Our interest focuses on determining the effective thermal conductivity of such a structure as a function of d, the degree of alignment.

The formulation introduced by Hao \textit{et al.} amounts to stating that, at the boundaries, the deviation of the phonon distribution from the 
local equilibrium is periodic. Using the notations from Figure \ref{fig:periodic_struct_bc2}, this condition can be written as
\begin{equation}
\left\{
\begin{array}{l}
f_{1}^{+}-f^{eq}_{T_{1}}=f_{2}^{+}-f^{eq}_{T_{2}} \\
f_{1}^{-}-f^{eq}_{T_{1}}=f_{2}^{-}-f^{eq}_{T_{2}}
\end{array}
\right.
\label{eq:periodic_conditions}
\end{equation}
where $f^{eq}_{T_1}$ and $f^{eq}_{T_2}$ refer to 
the equilibrium distributions at temperatures $T_1$ and $T_2$, 
where the superscript $+$ denotes particles moving to the right (with respect to figure \ref{fig:periodic_struct_bc2}) 
and where superscript $-$ refers to particles moving to the left.
This formulation enforces at the same time the periodicity of the heat flux and a temperature gradient.
In terms of deviational energy distributions, this relation becomes
\begin{equation}
\left\{
\begin{array}{l}
\hbar \omega(f_{1}^{+}-f^{eq}_{T_{eq}}-f^{eq}_{T_{1}})=\hbar \omega(f_{2}^{+}-f^{eq}_{T_{eq}}-f^{eq}_{T_{2}}) \\
\hbar \omega(f_{1}^{-}-f^{eq}_{T_{eq}}-f^{eq}_{T_{1}})=\hbar \omega(f_{2}^{-}-f^{eq}_{T_{eq}}-f^{eq}_{T_{2}})
\end{array}
\right.
\label{eq:periodic_conditions_dev1}
\end{equation}
which amounts to
\begin{equation}
\left\{
\begin{array}{l}
e^{d,+}_{1}-e^{eq}_{T_{1}}=e^{d,+}_{2}-e^{eq}_{T_{2}} \\
e^{d,-}_{1}-e^{eq}_{T_{1}}=e^{d,-}_{2}-e^{eq}_{T_{2}}
\end{array}
\right.
\label{eq:periodic_conditions_dev2}
\end{equation}

Computationally, this formulation can be implemented by emitting new particles from both sides while periodically advecting the existing particles. 
Without any loss of generality, let us assume that $T_1>T_2$. Particles emitted from the hot side originate from the distribution
\begin{equation}
e^{d,+}_{1}=e^{d,+}_{2}+e^{eq}_{T_{1}}-e^{eq}_{T_{2}}
\end{equation}
Therefore, at a given point on the boundary, denoting $\theta$ the angle with respect to the 
normal and $\phi$ the azimuthal angle, the flux per unit radial frequency locally emitted from boundary 1 (``hot'' side) 
in the solid angle $d \Omega=\sin \theta d \theta d \phi$ can be expressed as
\begin{align}
q''_{\omega,h}&= \sum_p e^{d,+}_{1}(\omega,\theta,\phi,p)\frac{D(\omega,p)}{4 \pi} V_{g}(\omega,p) \cos \theta \sin \theta d \theta d \phi \nonumber \\
&= \sum_p \underbrace{e^{d,+}_{2}\frac{D(\omega,p)}{4 \pi} V_{g}(\omega,p) \cos \theta \sin \theta d \theta d \phi}_{\rm crossing \ boundary \ 2} +\underbrace{(e^{eq}_{T_{1}}-e^{eq}_{T_{2}})\frac{D(\omega,p)}{4 \pi} V_{g}(\omega,p) \cos \theta \sin \theta d \theta d \phi}_{\rm new \ particles \ generated}
\end{align}
Similarly, the flux per unit radial frequency locally emitted from boundary 2 (``cold'' boundary) can be expressed as
\begin{equation}
q''_{\omega,c}= \sum_p \underbrace{e^{d,-}_{1}\frac{D(\omega,p)}{4 \pi} V_{g}(\omega,p) \cos \theta \sin \theta d \theta d \phi}_{\rm crossing \ boundary \ 1} -\underbrace{(e^{eq}_{T_{1}}-e^{eq}_{T_{2}})\frac{D(\omega,p)}{4 \pi} V_{g}(\omega,p) \cos \theta \sin \theta d \theta d \phi}_{\rm new \ particles \ generated}
\end{equation}
Hence the boundary condition can be enforced by:
\begin{itemize}
\item[i] Moving all particles and applying periodic boundary conditions to those crossing a periodic boundary: a particle leaving the system on one side is reinserted on the other side.
\item[ii] Generating new particles from the distribution
\begin{equation}
(e^{eq}_{T_{1}}-e^{eq}_{T_{2}})\frac{D(\omega,p)}{4 \pi} V_{g}(\omega,p)
\label{eq:particle_wall}
\end{equation}
The number of new particles is given by integrating (\ref{eq:particle_wall}) over all frequencies and polarizations and by multiplying the result by $\pi$
to account for the integration over the solid angle $\int_{\phi=0}^{2\pi} \int_{\theta=0}^{\pi/2} \cos \theta \sin \theta d\theta d \phi$.
The traveling direction of these particles is randomized on the half-sphere pointing into the domain and in the case of the hot boundary they are sent traveling to the right with a positive sign. 
Taking their mirror image, negative particles with the same properties are emitted by the cold boundary. 
\end{itemize}

\section{Validation}
\label{sec:validation}
\subsection{A ballistic problem}

In order to validate the proposed formulation, we first consider a one-dimensional system bounded by two isothermal 
(\ref{iso}) boundaries that are sufficiently close -- their distance, $L$, is much smaller than all phonon 
mean free paths -- that transport can be modeled as ballistic. The 
system is initially at a uniform equilibrium temperature $T_{0}$, when at $t=0^+$ the temperature of the isothermal walls 
impulsively changes to $T_0\pm \Delta T$. 

\begin{figure}[htbp]
\begin{center}
\includegraphics[width=.6\textwidth]{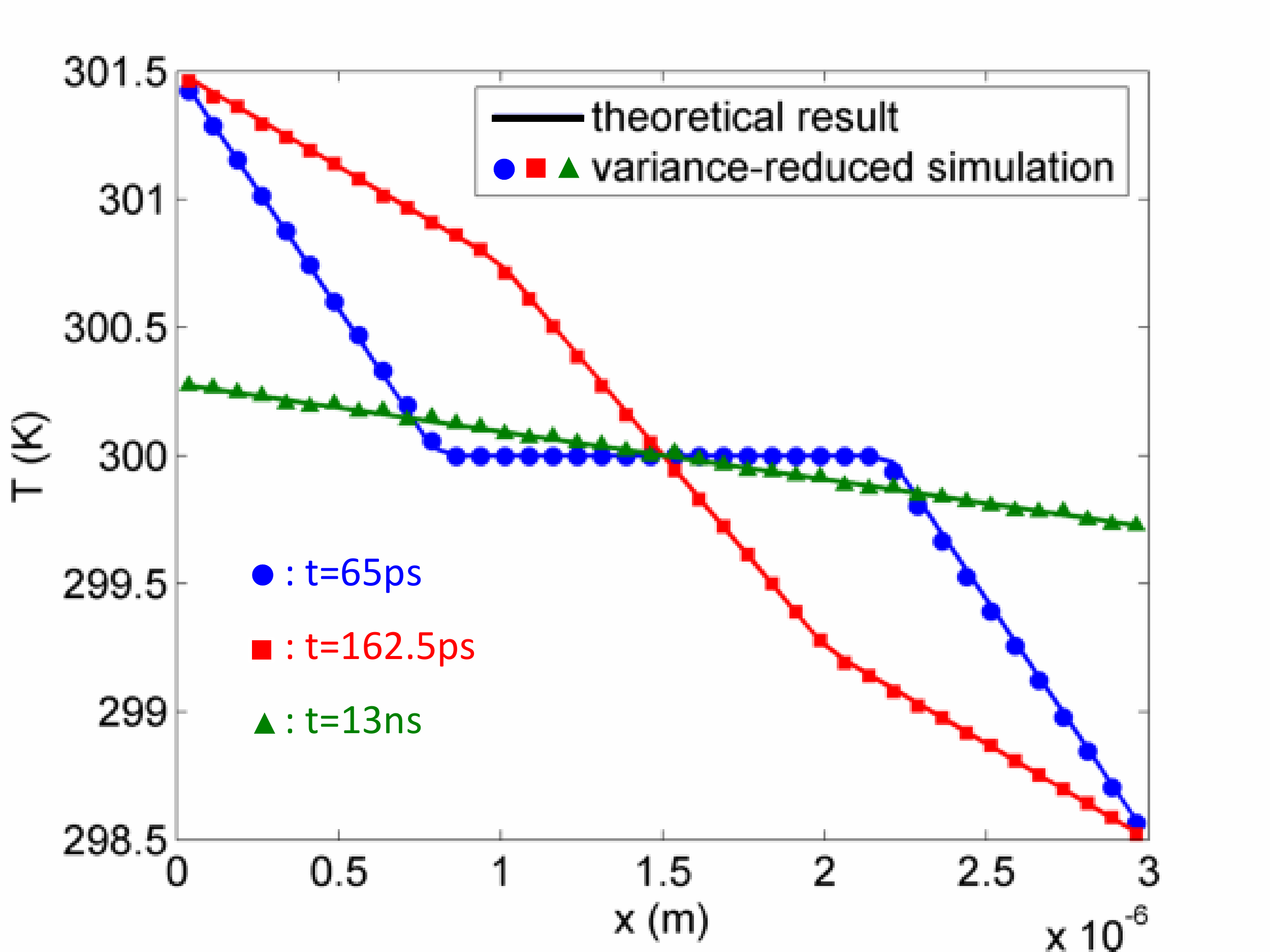}
\caption{Transient temperature profile in a one-dimensional ballistic system whose boundary temperatures undergo an 
impulsive change at $t=0$. Initially, 
the system is in equilibrium at temperature $T_{0}=300$K. At $t=0^+$, the wall temperatures become $T_{0}\pm \Delta T$; here, $\Delta T=3$ K. }
\label{fig:ballistic1}
\end{center}
\end{figure}

Appendix \ref{sec:ballistic} presents an analytical solution for the resulting transient evolution of the temperature field 
that is used here for comparison with our simulations. A particularly interesting case is the Debye model 
which, when coupled with small temperature amplitudes, 
allows a linearization of the general relation (\ref{eq:ballistic_breakdown}) to provide a fairly simple closed-form solution (\ref{eq:ballistic_closed_form}). 
Figure \ref{fig:ballistic1} shows a comparison between this solution and the variance-reduced Monte Carlo result.  
The simulation was run with $T_{eq}=T_0$ and the phonon velocity was taken to be  $12,360m.s^{-1}$ \cite{Hao09}. 
Excellent agreement is observed.

\subsection{Heat flux and thermal conductivity in a thin slab}
\label{subsec:thin_slab}
\begin{figure}[htbp]
	\centering
	\includegraphics[width=.5\textwidth]{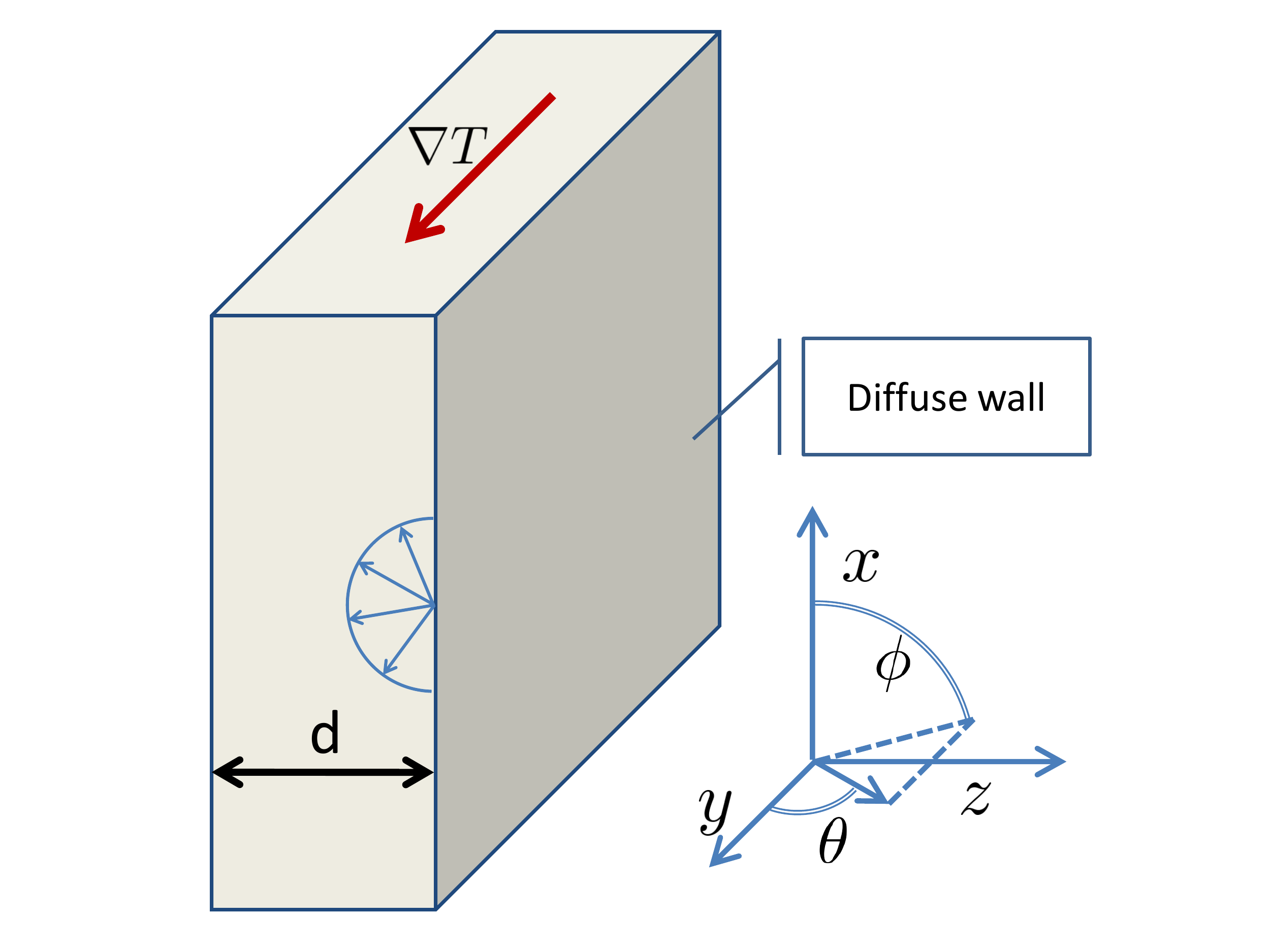}
	\caption{Heat conduction in a silicon slab due to an imposed temperature gradient in the $y$ direction. 
Slab is infinite in the $x$ and $y$ directions.}
	\label{fig:thin_film2}
\end{figure}

In this section we continue to validate our formulation by calculating the thermal conductivity of a thin silicon slab bounded by two diffusely reflecting walls a distance $d$ 
apart in the $z$ direction (see Figure \ref{fig:thin_film2}). The slab is infinite in the $x$ and $y$ directions.

\begin{figure}[htbp]
	\centering
	\includegraphics[width=.5\textwidth]{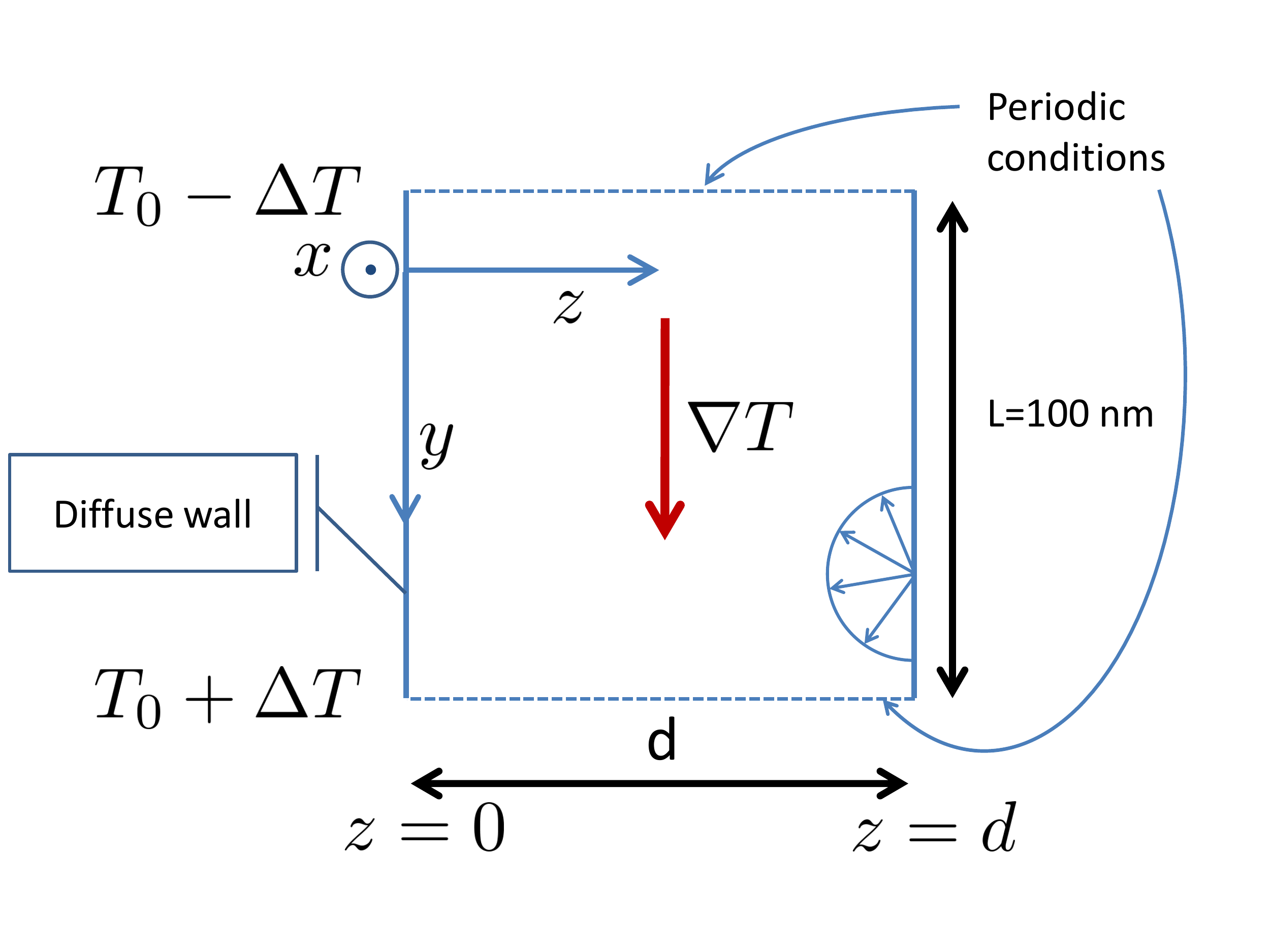}
	\caption{Simulation geometry. Boundaries at $z=0$ and $z=L$ are diffusely reflecting. Infinite domain in 
$y$ direction is terminated using periodic boundaries $L=100$nm apart.}
	\label{fig:thin_film}
\end{figure}

This problem is considered here because the solution can be expressed analytically. We introduce the local deviation 
function $f^d=f-f^{loc}$ and, denoting the temperature gradient by $dT/dy$, rewrite the BTE at steady state as:
\begin{equation}
V_g \frac{d f^{loc}}{dT} \frac{dT}{dy} \cos(\theta) + \mathbf{V}_g \nabla f^d = -\frac{f^d}{\tau}
\end{equation}

This equation can be solved to yield, in the coordinate system introduced in Figure \ref{fig:thin_film2}, 
\begin{align}
&f^{d}(z,\omega,p,\theta,0<\phi<\pi)= \nonumber \\
&-\Lambda(\omega,p,T_{0}) \cos(\theta) \frac{df^{loc}(\omega,T_{0})}{dT} \frac{dT}{dy}\left\{1-\exp \left[-\frac{z}{\Lambda(\omega,p,T_{0}) \sin(\theta) \sin(\phi)} \right] \right\} \label{eq:phonon_slab1} \\
&f^{d}(z,\omega,p,\theta,-\pi<\phi<0)= \nonumber \\
&-\Lambda(\omega,p,T_{0}) \cos(\theta) \frac{df^{loc}(\omega,T_{0})}{dT} \frac{dT}{dy}\left\{1-\exp \left[-\frac{z-d}{\Lambda(\omega,p,T_{0}) \sin(\theta) \sin(\phi)} \right] \right\} \label{eq:phonon_slab2}
\end{align}
where $\Lambda(\omega,p,T_{0})$ is the average mean free path at frequency $\omega$, polarization $p$ and temperature $T_0$, given by
\begin{equation}
\Lambda(\omega,p,T_{0})=V_g(\omega,p)\tau(\omega,p,T_0)
\end{equation}
Moments of this solution can be numerically integrated to yield values for the heat flux and the thermal conductivity of the slab. 

In the simulation, we calculate the thermal conductivity by measuring the steady state heat flux in response to a 
temperature gradient along the $y$ axis (see Figure \ref{fig:thin_film}). Due to the translational 
symmetry of the system, we impose the temperature gradient using the periodic unit-cell formulation presented in 
section \ref{subsubsec:periodic_boundaries}, which allows us 
to use a finite system size in the $y$-direction, taken to be $L=100$nm. In order to measure the thermal conductivity at 
$T_0$, a temperature gradient is imposed by setting a target temperature of $T_{0}+\Delta T$ for the hotter of the two 
boundaries and $T_{0}-\Delta T$ for the colder boundary, and we
proceed as explained in \ref{subsubsec:periodic_boundaries}. The deviational method allows the solution of this problem 
for $\Delta T\ll T_0$ (here, $\Delta T=0.05$K), in contrast to 
non-variance-reduced methods that would require $\Delta T\sim T_0$ to achieve statistically significant results. 
The best choice for the equilibrium (control) temperature is clearly $T_{eq}=T_0=300$K. 
Initializing the simulation at equilibrium at $T_0$ is also convenient, because no particles need to be generated for the initial configuration.

Figure \ref{fig:thin_film_valid}  compares the heat flux in the $y$ direction inside a slab of silicon 
(see Appendix A for material parameters) of thickness $d=$100nm, as computed by the deviational method, 
to the analytical solution. Figure \ref{fig:thin_film_valid2} compares the thermal conductivity of the slab at $T_0$=300K as a function of $d$ 
computed from the deviational method and from the analytical expression. Very good agreement is observed in all cases.

\begin{figure}[htbp]
\begin{center}
\includegraphics[width=.8\textwidth]{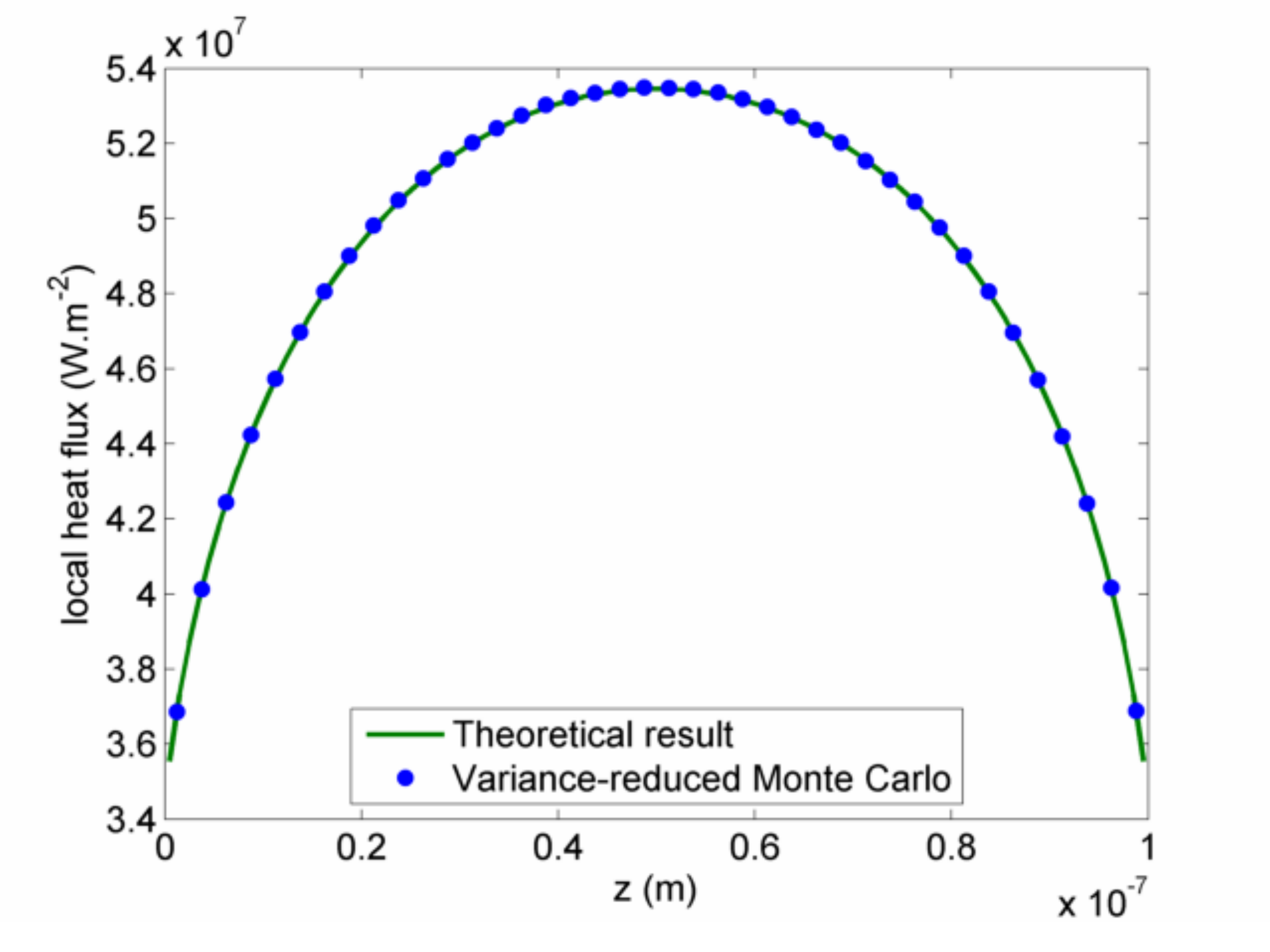}
\caption{Spatial variation of the axial (in the $y$ direction) heat flux in a thin film with a thickness $d=100$nm, 
computed theoretically and compared to the result of the deviational simulation. }
\label{fig:thin_film_valid}
\end{center}
\end{figure}

\begin{figure}[htbp]
\begin{center}
\includegraphics[width=.8\textwidth]{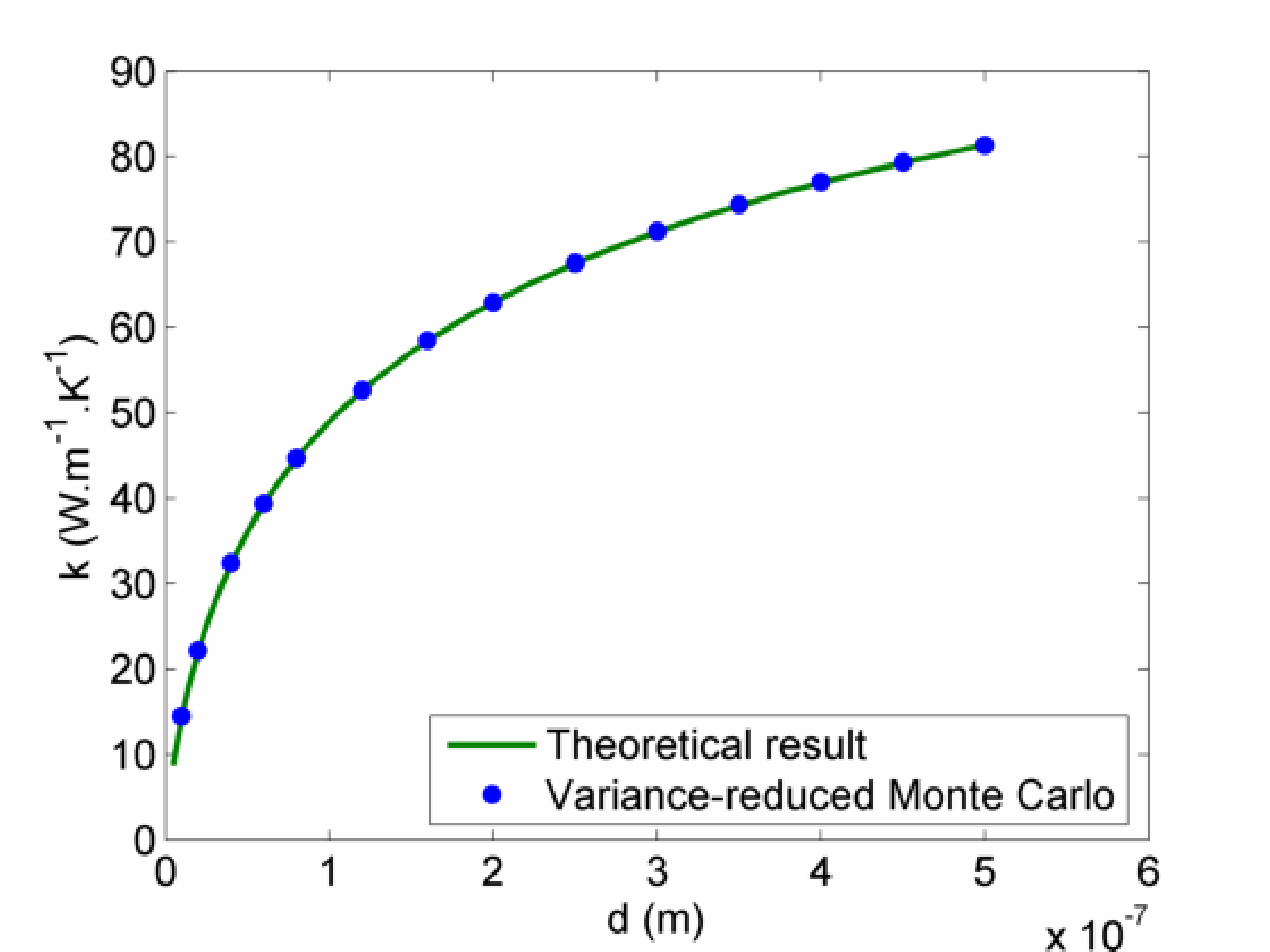}
\caption{Theoretical values of the thin film thermal conductivity at $T_0=300$K, computed by numerical integration of the 
theoretical expressions  (\ref{eq:phonon_slab1}) and (\ref{eq:phonon_slab2}). Comparison with the values obtained from the deviational simulation.}
\label{fig:thin_film_valid2}
\end{center}
\end{figure}

\section{Computational efficiency}
\label{sec:efficiency}
The variance-reduced method developed here allows substantial improvement in the relative statistical uncertainty, 
$\sigma/\Delta T$, compared to non-variance-reduced simulations. Here, $\sigma$ is the standard deviation in the 
temperature measurement and $\Delta T$ is the characteristic 
temperature difference (as, for example, in the validation case studied in \ref{subsec:thin_slab}). 

\begin{figure}[htbp]
\begin{center}
\includegraphics[width=.6\textwidth]{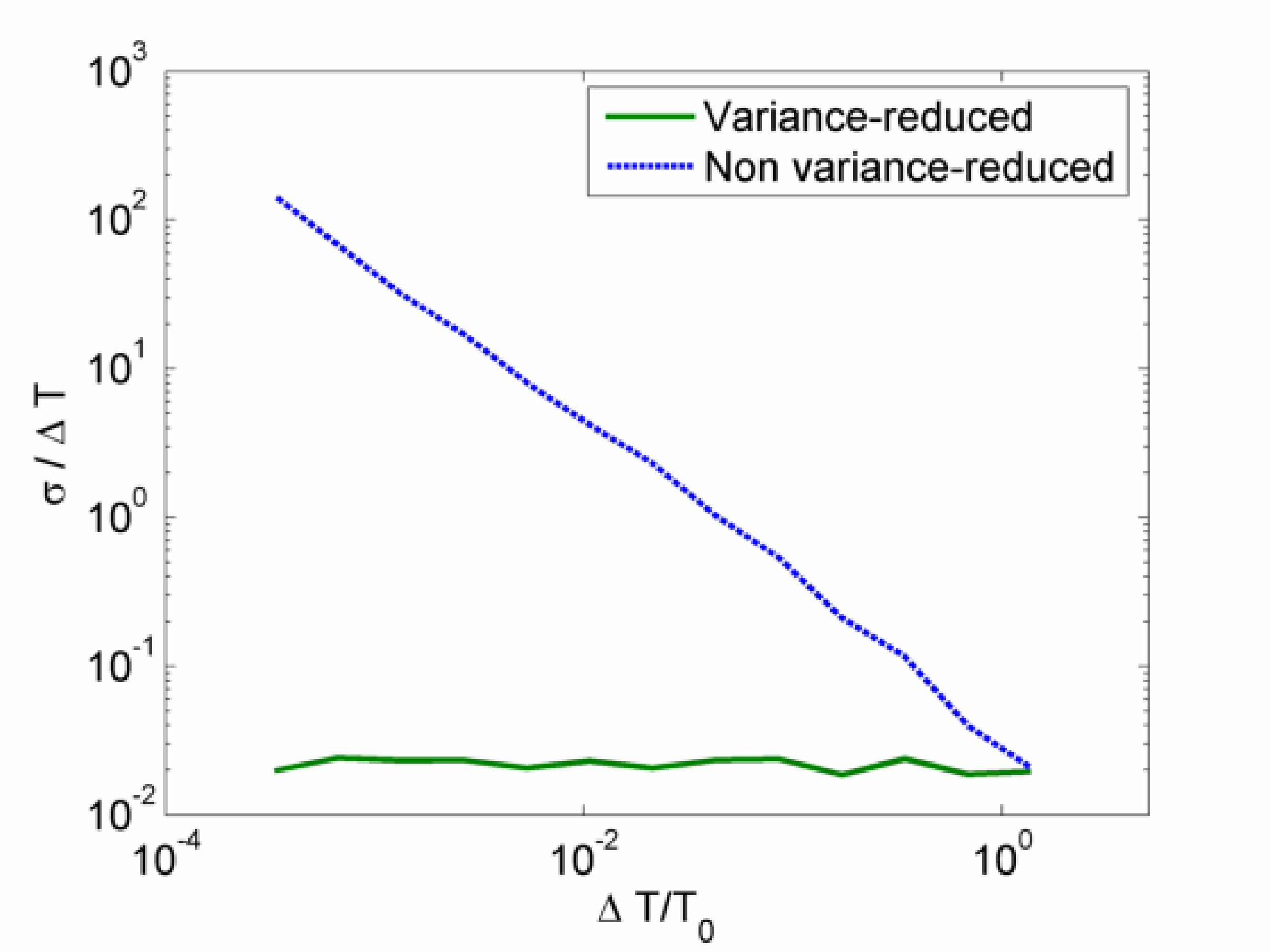}
\caption{Comparison of relative statistical uncertainties for equilibrium systems at temperature $T_{1}$ with $\Delta T=T_1-T_{0}$ and $T_{0}=300K$.}
\label{fig:variance}
\end{center}
\end{figure}

Figure \ref{fig:variance} compares the relative statistical uncertainty of the variance-reduced 
with the standard method. The reported data was obtained by simulating equilibrium at some temperature $T_1$, 
and defining $\Delta T=T_1-T_0$ as the characteristic signal that needs to be resolved.
By choosing $T_{eq}=T_0$ in the deviational method, we ensure finite deviation from equilibrium is considered and thus 
the statistical uncertainty is non-zero. Simulating an equilibrium state is a matter of convenience, 
because in non-equilibrium problems the number of particles and thus the local 
statistical uncertainty varies as a function of space 
in the deviational simulation and is thus difficult to quantify precisely; simulations of simple problems 
(e.g. Couette-type problems) in the past \cite{Baker05,homolleb07,radtke09} have yielded very similar results. 
We also note that even though $\sigma/\Delta T$  is strictly speaking the ratio of statistical uncertainties, 
it serves as a good approximation to the ratio of computational cost, because the cost of the deviational simulation 
is very similar to that of standard Monte Carlo methods. Specifically, the speedup provided by the deviational method 
is given by the {\bf square} of the relative statistical uncertainties.

A very interesting feature of  variance-reduced methods is that 
the standard deviation of the results is proportional to the amplitude $\Delta T$ of the signal, 
as shown in Figure \ref{fig:variance} (see also \cite{Baker05,Hadjiconstantinou06,Hadjiconstantinou10}). 
As a consequence, variance-reduced methods are able to provide the desired relative statistical uncertainty (signal to noise ratio) for arbitrarily low signals 
without requiring more computational effort. In contrast, in the case of the non-variance-reduced method, 
it is more computationally expensive to obtain the desired level 
of relative statistical uncertainty for small variations in temperature, than for large variations in temperature. 
In these methods, for $\Delta T<<T_0$,  the statistical 
uncertainty is approximately constant (set by equilibrium fluctuations) and thus $\sigma/\Delta T \sim 1/\Delta T$. 
As a result, the speedup offered by the variance-reduced methods scales as $1/(\Delta T)^2$. For example 
at $\Delta T/T_0 \approx 10^{-2}$ (i.e. $\Delta T \approx 3K$ at room temperature) the speedup is approximately 
4 orders of magnitude (see Figure \ref{fig:variance}); at $\Delta T/T_0 \approx 10^{-3}$, the speedup is approximately 
6 orders of magnitude.

\section{Applications}
In this section we present some applications of the deviational method to problems of current engineering interest. Modeling work in these areas 
is still ongoing; the objective of this discussion is mainly to showcase the capabilities of the proposed method.
\subsection{Thermal conductivity of nanoporous silicon: influence of nanopore alignment}
\label{subsec:staggered-aligned}
\begin{figure}[htbp]
\begin{center}
\includegraphics[width=.8\textwidth]{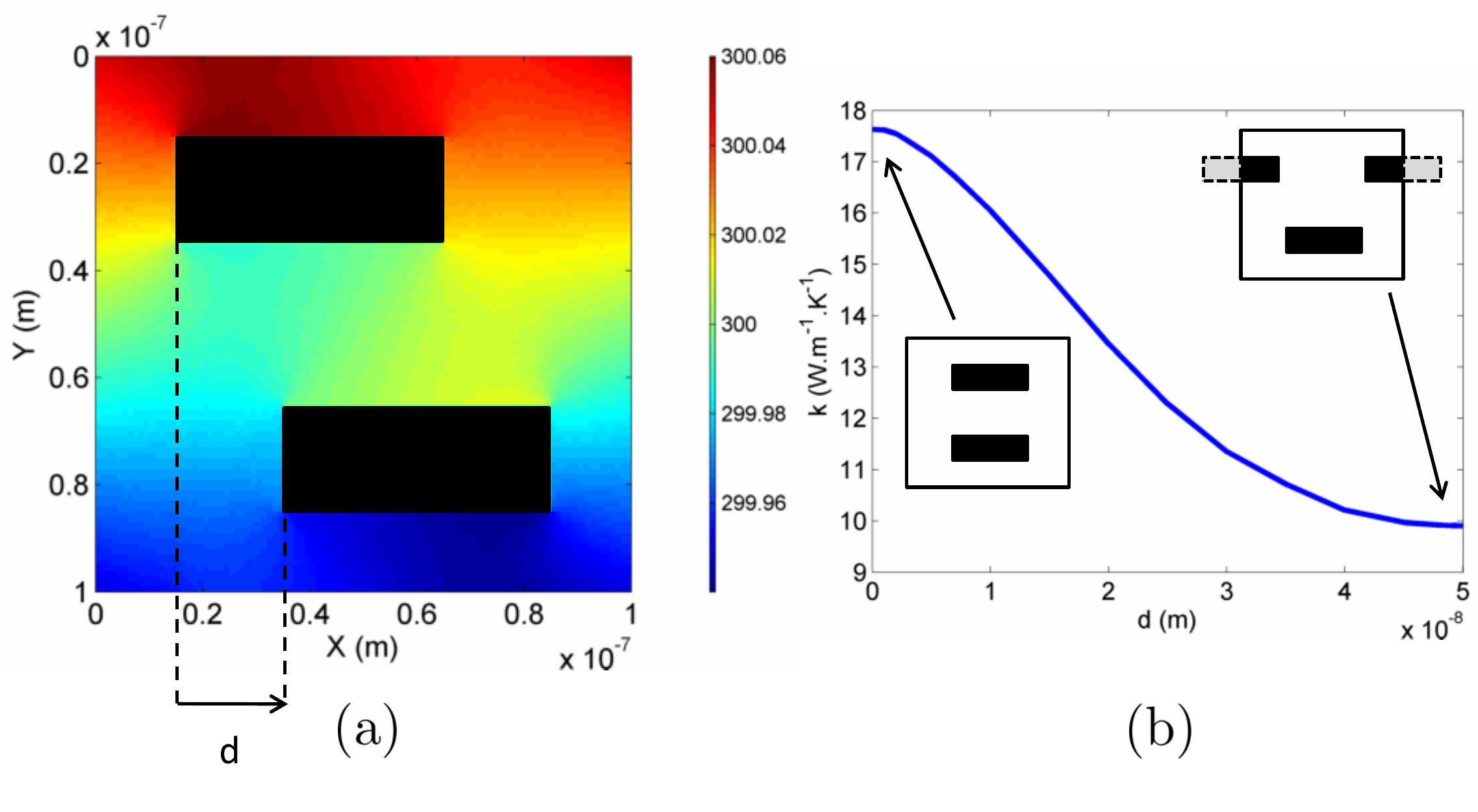}
\caption{(a)  Temperature field in a unit cell of a periodic nanoporous material. 
(b) Thermal conductivity as a function of parameter d.}
\label{fig:rectangles_quincunx}
\end{center}
\end{figure}

Decreasing thermal conductivity as a means of improving the thermoelectric effect has received considerable attention, 
and nanostructures are a novel approach towards this goal. 
Similarly to Huang \textit{et al.} \cite{Huang09} and Jeng \textit{et al.} \cite{Jeng08}, we assess here the thermal conductivity of novel nanostructured materials. 
The nanostructure considered here is made of rectangular pores as shown in Figure \ref{fig:periodic_struct_bc2}. 
We model it as a 2D problem (possible 
if the material boundaries in the directions normal to the plane shown in the figure can be approximated as specularly reflecting). 
Figure \ref{fig:rectangles_quincunx} shows the periodic cell considered and defines the parameter d that we use to describe the spatial distribution of the 
pores. The thermal conductivity in the $y$ direction is measured by imposing periodic unit-cell boundary conditions 
as explained in section \ref{subsubsec:periodic_boundaries} with a 
temperature difference of 0.1K across the unit cell. Using the data of Appendix A, the contributions of the different 
mean free paths to the bulk thermal conductivity can be calculated. 
A plot of the effective thermal conductivity, as computed with the deviational variance-reduced method, 
is displayed in Figure \ref{fig:rectangles_quincunx}.  The thermal conductivity is reduced by almost a factor of 2 because
 of this geometrical effect. This highlights the importance of ballistic effects.

\begin{figure}[htbp]
\begin{center}
\includegraphics[width=.8\textwidth]{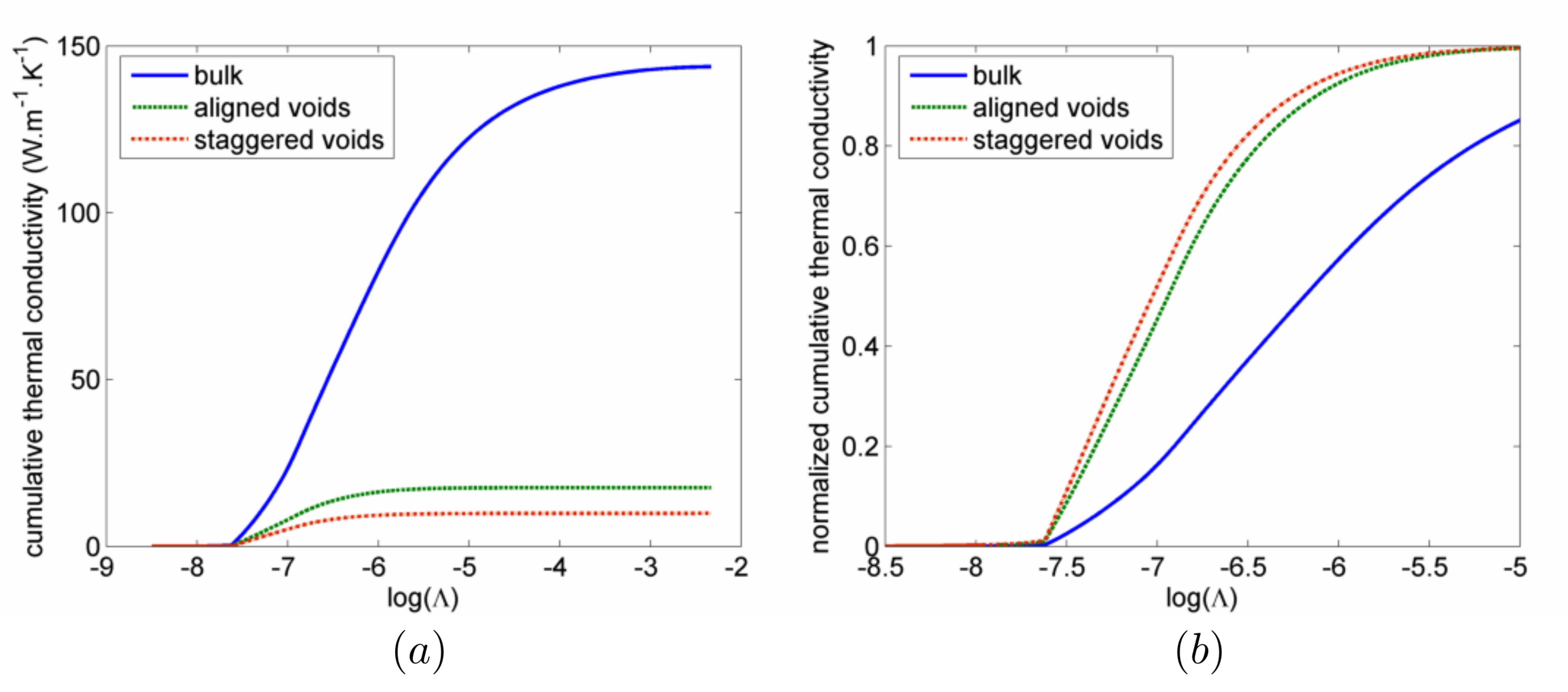}
\caption{(a) Thermal conductivity accumulation as a function of the mean free path. 
The bulk conductivity was computed by numerical integration of the thermal conductivity per unit frequency 
$\tau v^{2}C_{\omega}/3$ \cite{minnich,JPThesis, chenbook}; here, $C_{\omega}$ is the heat capacity per frequency unit. 
(b) Normalized thermal conductivity accumulation, highlighting the influence of ballistic effects on the thermal conductivity.}
\label{fig:cumulated_thermal_cond}
\end{center}
\end{figure}

The importance of ballistic effects is further highlighted by Figure \ref{fig:cumulated_thermal_cond} which shows that
 at $T_0=300$K, mean free 
paths from 50nm to 10$\mu$m contribute significantly to the thermal conductivity of the bulk material;
the presence of voids with period of 100nm affects the contribution of all 
mean free paths, but completely suppresses the contribution of all mean free paths greater than about one micrometer. 
Tuning the alignment parameter, decreases further the contribution of the mean free paths between 50nm and $1\mu$m. 

\subsection{Simulation of thermal conductivity spectroscopy}
\label{subsec:laser_pulse}
\begin{figure}[htbp]
\begin{center}
\includegraphics[width=.8\textwidth]{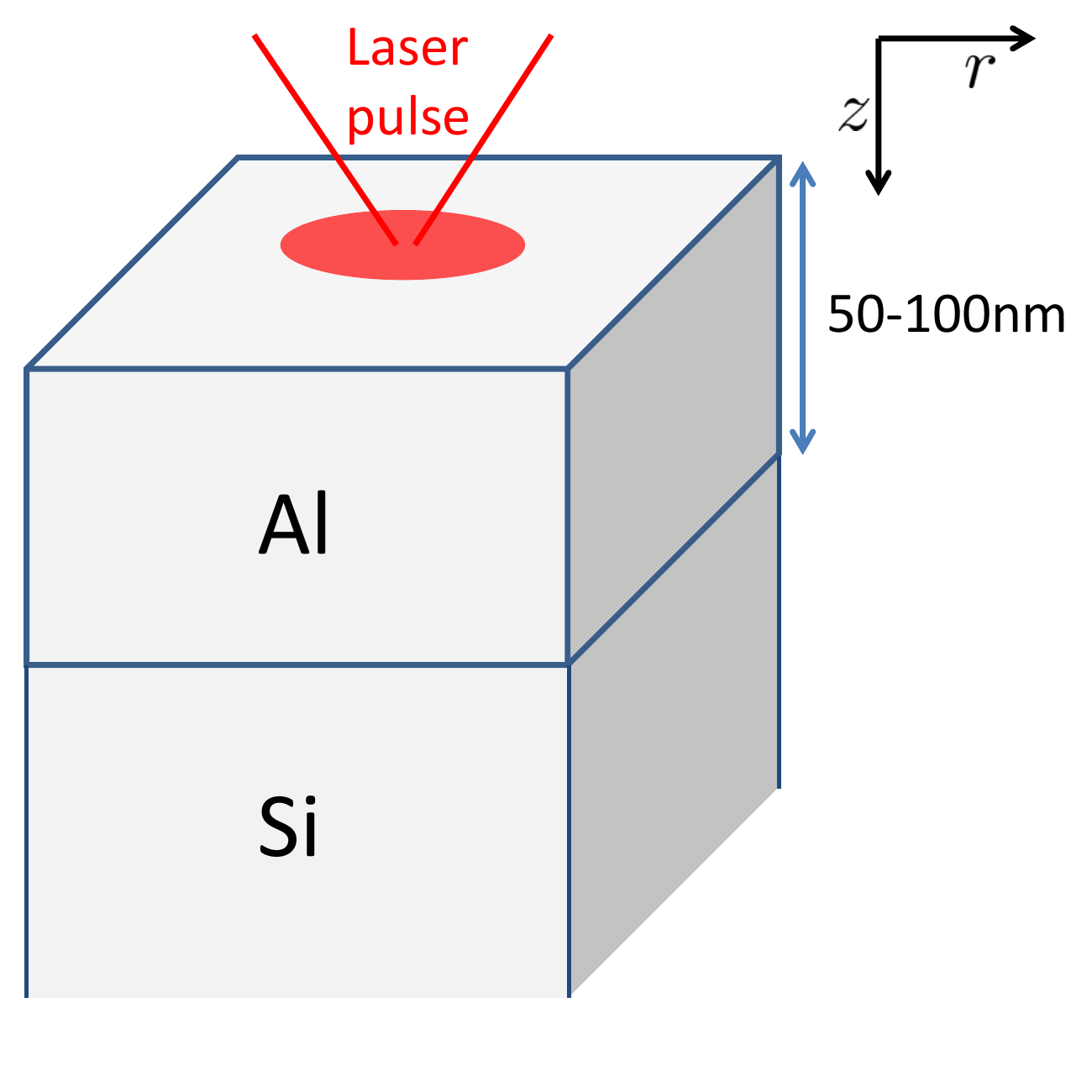}
\caption{System composed of a slab of aluminum on a semi-infinite silicon wafer, used for transient thermoreflectance 
(TTR) experiments. At t=0, a laser pulse induces a temperature field $T(r,z,t)$. The temperature field evolution after the pulse is computed by assuming that the aluminum surface is adiabatic.}
\label{fig:expe_setup}
\end{center}
\end{figure}

Figure \ref{fig:expe_setup} depicts an experimental setup developed in the MIT Nanoengineering Lab \cite{minnich11} as a prototype ``thermal conductivity spectroscopy'' system. This experiment is based on pump-probe transient thermoreflectance, in which a pump pulse is used to change the physical properties of a sample and a probe pulse is used to measure the change. 
In this experiment, a thin film of aluminum (thickness between 50 and 100nm) is deposited on a silicon wafer 
and is initially at uniform temperature, say 300K. At $t=0$, localized laser irradiation creates a hot spot, 
shown in figure \ref{fig:expe_setup} as centered on the origin $(r=0,z=0)$ 
of the coordinate system. A reliable description of the subsequent evolution of the temperature field is central to interpreting the experimental 
results and creating a means for inferring phonon mean free paths (the goal of this experiment) from experimental measurements (e.g. surface temperature)

Given the scale of the aluminum slab, the impulsive nature of the heating, and the short duration of the 
phenomenon, phonon ballistic behavior needs to be accounted for, 
necessitating a Boltzmann treatment. However, this problem is very difficult (if not impossible) to simulate using standard Monte Carlo methods: 
the initial perturbation to the temperature field is small in amplitude (see below) which makes resolution of transient 
results very costly. Moreover, the need to simulate early as well as 
late times and avoid artifacts from artificial domain termination makes the simulation of a large computational domain necessary, even though the original hot spot 
is very small. In traditional Monte Carlo methods, this large computational domain would need to be filled with particles. 

The method proposed makes this calculation possible. Simulating the deviation from equilibrium allows the calculation 
to proceed using zero particles in regions 
not yet affected by the heating pulse.  Thus, in addition to variance reduction which removes the limitations 
associated with statistical uncertainty, 
simulating the deviation from equilibrium simultaneously considerably reduces the computational cost resulting 
from the multiscale nature of this problem. We also note that by taking the equilibrium distribution at 300K, 
the simulation only has positive particles. Hence there will be no cancellation of particles and the entire simulation 
will run with a fixed amount of particles. 

In practice, one can exploit the cylindrical symmetry in order to reduce the problem dimensionality: the resulting temperature field is expected to depend only on 
the depth $z$ and on the distance from the center of the pulse, $r$. Therefore, we can use toroidal cells to sample the temperature and process the scattering. The 
only drawback is that cells near the center, at small radius, will have a smaller volume and will sample the temperature over a smaller number of particles, thus 
yielding noisier results in these regions.

\subsubsection{Initial condition}
As stated above, since the material is originally at equilibrium at $T_0=300$K it is most convenient, but also computationally 
efficient, to choose $T_{eq}=T_0$. Laser irradiation introduces a heating effect 
in a thin layer close to the irradiated surface which has been parametrized \cite{minnich} using the 
following expression
\begin{equation}
\Delta T(r,z)= T-T_0=A\exp \left(-\frac{2r^2}{R_0^{2}}-\beta z \right)
\end{equation}
with $A=1$K, $R_0=15\mu$m and $\beta^{-1}=7$nm. This expression is used here as an initial condition for the material temperature. 
Regions for which $\Delta T < .005K$ were taken to be at equilibrium at  $T_0=T_{eq}$ (no particles). 

\subsubsection{Interface modeling}
\label{subsubsec:interface_modeling}
The top surface of the aluminum material ($z=0$) is modeled as a diffusely reflecting wall.
 
Modeling the interface between the two materials accurately is still an active area of research. Here, we chose to use 
a recently developed model \cite{minnich,Minnich11_not_submitted} which 
relates the transmissivity to the interface conductance $G$  through the expression
\begin{equation}
<P_{1\rightarrow 2}C_1V_{g,1}>=\frac{2}{\frac{1}{<C_1V_{g,1}>}+\frac{1}{<C_2V_{g,2}>}+\frac{1}{2G}}
\label{eq:interface_probability}
\end{equation}
Here, $P_{i\rightarrow j}$ denotes the probability for a phonon to pass through the interface from material $i$ to $j$; 
the brackets denote integration over frequency and sum over polarization, while $C_{1}$ and $C_{2}$ denote the 
volume heat capacity per unit frequency in media 1 and 2, respectively. In this model, we assume 
that the interface is totally diffuse: the direction of an incident particle is reset 
regardless of the transmission or reflection of the particle, while its frequency and polarization are retained 
\cite{Chen98}. 
For the interface conductance G, we use the experimental value $G=1.1 \times 10^{8} \mathrm{W m^{-2}K^{-1}}$ 
\cite{Minnich11_not_submitted}.

We also utilize the expression \cite{Chen98}
\begin{equation}
D_1(\omega,p)V_{1,g}(\omega,p)f^{eq}_{T_0}P_{1\rightarrow 2}(\omega,p)=D_2(\omega,p)V_{2,g}(\omega,p)f^{eq}_{T_0}P_{2\rightarrow 1}(\omega,p)
\label{eq:equilibrium_interface}
\end{equation}
which relates the probability for a phonon with radial 
frequency $\omega$ and polarization p to pass 
through the interface from 1 to 2  to the probability to pass from 2 to 1.

We can easily verify that relation (\ref{eq:equilibrium_interface}) applies when the deviational energy $e^{d}$ 
is used instead of the phonon distribution. Additionally, expression (\ref{eq:interface_probability}) which relies,
among other things, on (\ref{eq:equilibrium_interface}) \cite{minnich,Minnich11_not_submitted}, also remains unchanged when applied to
deviational particles.

Following \cite{minnich,Minnich11_not_submitted}, we let $P_{1\rightarrow 2}$ be a constant (which makes it easy to calculate from 
(\ref{eq:interface_probability})) and deduce $P_{2\rightarrow 1}$ from (\ref{eq:equilibrium_interface}). 
In our case we chose to set $P_{Al\rightarrow Si}$ constant, except for the high frequency transverse acoustic modes; 
 since the cutoff frequency of the TA branch in Si is lower than the TA cutoff frequency in Al, phonons with 
such frequencies must undergo total reflection \cite{minnich}. Similarly, LA phonons in Si whose frequency is above the 
aluminum LA branch cutoff frequency are totally reflected. 

\subsubsection{Domain termination}
\label{subsubsec:domain_termination}
At long times, phonons may travel far from the hot spot. In order to avoid discretizing an infinite domain with 
computational cells (for calculating the temperature) we restrict our discretization to a finite (but large) 
``nominal'' domain. In order to simulate accurately and consistently 
the actual system, we keep track of the particles even after they have left the nominal part of the domain. 

Particles that leave this domain are not sampled 
(for calculating the temperature and pseudo temperature), but are still scattered by assuming a local temperature 
of 300K as an input parameter for the relaxation time. This amounts to a linearization of the collision operator at $T=300$K 
and is based on the reasonable assumption that sufficiently 
far from the heating source, the temperature is very close to 300K. Particles that leave the nominal part of the domain 
may reenter the nominal domain, hence ensuring a rigorous treatment of the semi-infinite region. 

Particular care is taken to ensure that the frequency and 
polarization of a particle is drawn from the correct distribution, because energy conservation---built into the 
simulation method---requires that the number of particles is conserved by the scattering process 
and is inconsistent with approximations which do not conserve energy. 
For example, setting $T_{loc}=$300K  is inconsistent with energy conservation because $e^{loc}(T_{loc}=300K)-e^{eq}=0$, 
which implies no particle generation, which in the presence of particle deletion due to the term $-e^d/\tau$ leads 
to net particle and thus energy loss. This situation can be rectified by allowing the 
temperature at the particle position to be different from $T_{eq}$; specifically, we write $T=T_{eq}+\epsilon$ and expand 
\begin{equation}
\frac{D(\omega,p)(e^{loc}-e^{eq}_{T_{eq}})}{\tau(\omega,p,T_{eq})} \approx \frac{D(\omega,p)}{\tau(\omega,p,T_{eq})} \frac{\partial e^{eq}_{T_{eq}}}{\partial T} \epsilon
\label{eq:linear_approx}
\end{equation}
Frequencies and polarizations are thus drawn from
\begin{equation}
\frac{D(\omega,p)}{\tau(\omega,p,T_{eq})} \frac{\partial e^{eq}_{T_{eq}}}{\partial T}
\end{equation}
since (\ref{eq:linear_approx}) once normalized, does not depend on the local $\epsilon$. 
As before, energy conservation is ensured by simply conserving the particles. 

In addition to providing a method for terminating simulations, this approach represents a promising avenue for 
treating the entire simulation domain in the limit that linearization of the collision operator is 
appropriate. The advantage of this formulation is significant reduction in computational cost 
because evaluation of the local temperature and pseudo-temperature is not required every timestep. 
Further details will be presented in a future publication.

\subsubsection{Simulation results}
\begin{figure}[htbp]
\begin{center}
\includegraphics[width=.8\textwidth]{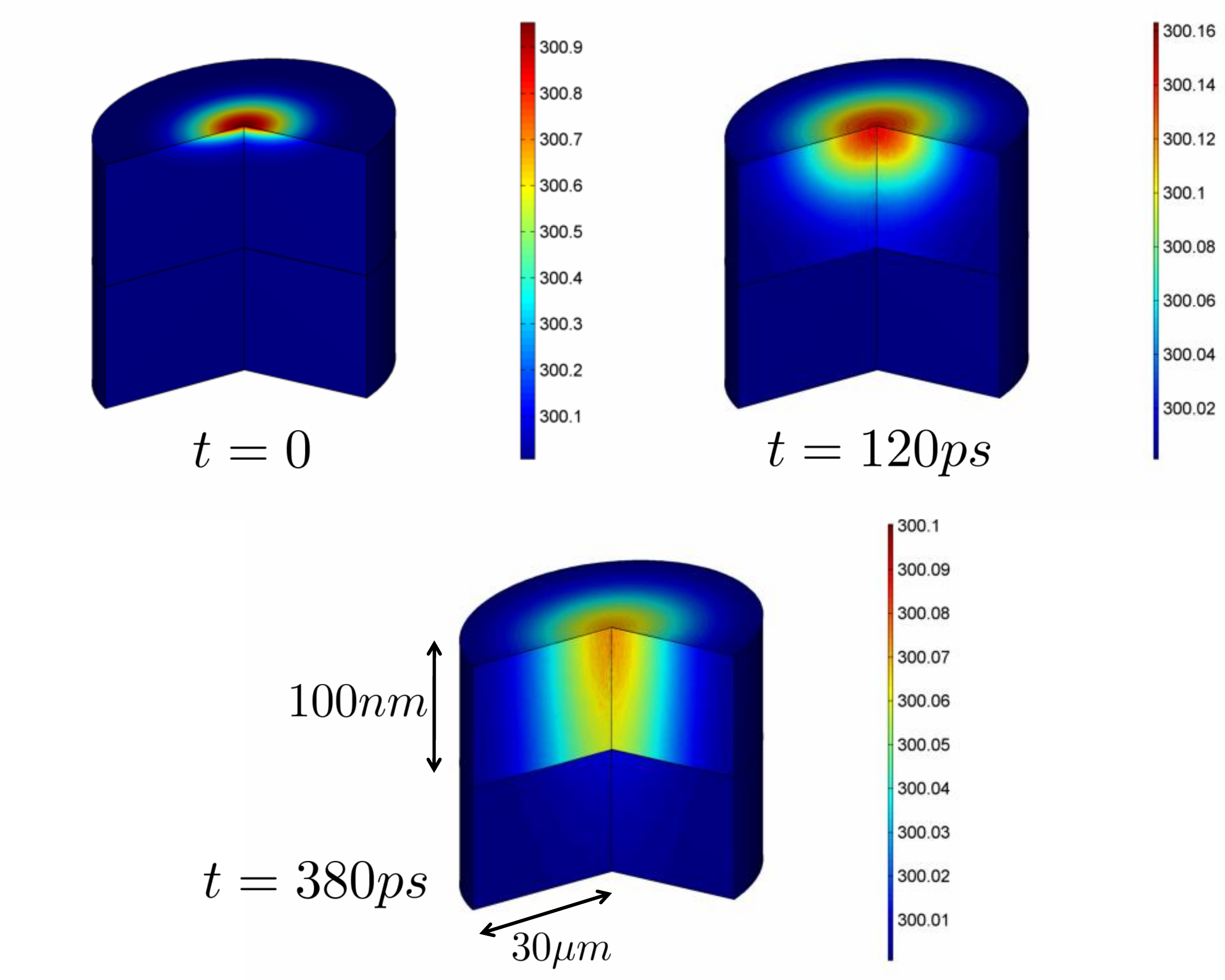}
\caption{Variance-reduced temperature field in  aluminum slab and the silicon wafer after initial heating by a laser pulse.  
The picture shows the aluminum slab (100nm thickness) and a portion of the silicon wafer 
(100nm thickness).}
\label{fig:phonon_results}
\end{center}
\end{figure}

\begin{figure}[htbp]
\begin{center}
\includegraphics[width=.8\textwidth]{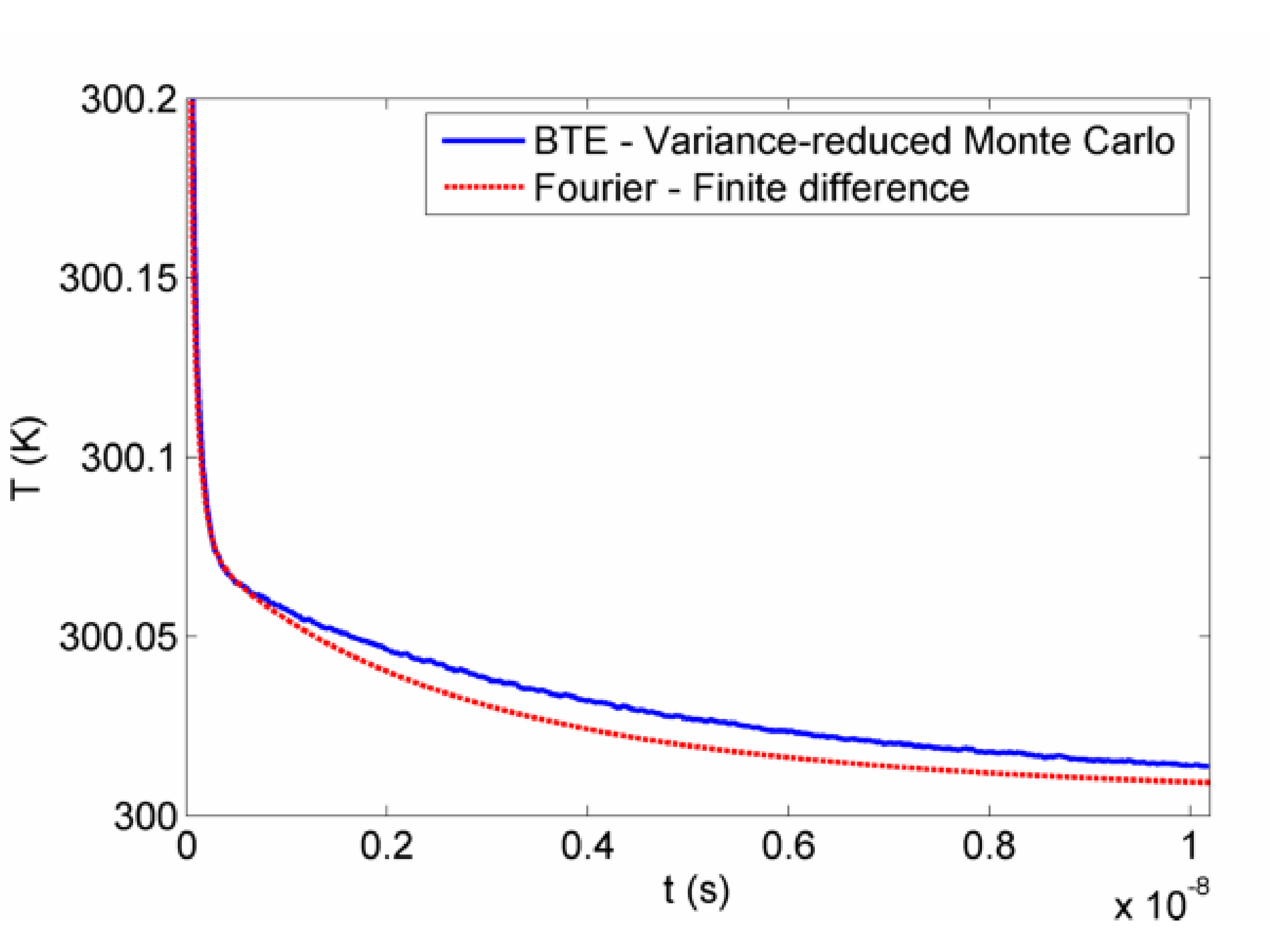}
\caption{Surface temperature at the hot spot (averaged over the region $0\leq r \leq 2\mu m$, $0\leq z\leq 5nm$) 
as a function of time after initial heating by laser pulse. The difference with the solution based on the Fourier model 
is a result of ballistic effects.}
\label{fig:surface_temperature}
\end{center}
\end{figure}

Figure \ref{fig:phonon_results} and \ref{fig:surface_temperature} show
that the
variance-reduced method developed here
can calculate the temperature field with small statistical uncertainty.
This is
remarkable given the minute temperature
differences ($O(0.01)$K) present in this problem, especially at late
times. For
such temperatures,
according to Figure \ref{fig:variance}, the
speedup compared to a standard Monte Carlo method is on the order of $10^9$.

Figure \ref{fig:surface_temperature} compares our simulation results with a
numerical solution of the heat conduction
equation (Fourier's Law). The differences between the two predictions are a result of non-diffusive
(ballistic/transitional) effects. The detailed information available in
simulations of this phenomenon can  assist in the development of methodologies
for characterizing carrier mean free paths from comparisons such as the one
shown in Figure 14.
Here, we note that the present calculation
does not account for thermal transport by electrons in aluminum. This was
neglected in the interest of
simplicity and because
the primary focus of this experiment is transport through the silicon
substrate
\cite{minnich}.
Thermal transport by electrons in aluminum will be considered and
evaluated in a
future publication.

\section{Discussion}

We have shown that efficient and accurate algorithms for solving the BTE with significantly reduced statistical uncertainty can 
be developed by focusing on the deviation from a nearby equilibrium within an energy-based formulation. The energy-based formulation  facilitates 
exact energy conservation thus improving the simulation fidelity, while the variance reduction is made possible by the deterministic information inherent in the Bose-Einstein 
distribution which describes the nearby equilibrium.  
The proposed method was validated  using analytical solutions of the 
Boltzmann Transport Equation. Very good agreement with the analytical results was found.

The proposed algorithm was used to study the effect of porosity on the effective thermal conductivity of pure silicon. Our results show that staggering 
periodically arranged voids at small scales exploits  ballistic shading to effect reduction in the effective thermal conductivity. A more 
systematic investigation of the effects of porosity on the effective conductivity of silicon---including anisotropic effects---will be the subject of future work. 

We also presented simulations of a recently developed experimental technique known as thermal conductivity spectroscopy, in which the transient response of a thin 
aluminum slab over a silicon wafer to a localized heating induced by a laser pulse is used to infer properties of 
heat carriers. This simulation required the development of a domain termination algorithm for rigorously treating 
deviational particles as they travel to regions far from the heating source, 
without having to sample these particles everywhere in this semi-infinite region. The algorithm developed corresponds 
to a linearization of the collision operator and may, in fact, form the basis of a significantly more efficient simulation 
approach valid in cases where linearization is appropriate.

In addition to illustrating the benefits of variance reduction, simulations of the thermal conductivity spectroscopy 
problem also showcase the value of  
the proposed simulation approach as a new multiscale method: in contrast to typical multiscale methods which focus on spatially decomposing the domain into the particle and continuum subdomains, 
the present algorithm achieves a seamless transition from one description to the other 
by instead {\it algebraically} decomposing the distribution function into a part described by particles and a part described deterministically \cite{submitted}. 
Although here the simplest such implementation has been presented (deterministic description is equilibrium at 
temperature $T_0\neq T_0(\mathbf{x},t)$), deviational algorithms featuring a deterministic description 
that varies as  a function of space ($e^{eq}=e^{eq}(\bf{x})$) have been developed \cite{homolleb07,radtke09} and shown to achieve improved variance reduction 
as $Kn\rightarrow 0$ \cite{radtke09}, albeit at the cost of a moderately more complex algorithm. 
In the problem considered here, the continuum behavior at large 
distances from the heat source is in fact equilibrium at $T_0$ and thus the present algorithm is sufficient. 
However, in other problems where a local equilibrium 
is present in large parts of the domain, algebraic decomposition using $e^{eq}=e^{eq}(\bf{x})$ will be able to provide considerable 
computational savings by considerably reducing the number of particles required for its simulation. 
    
\section{Acknowledgements}
The authors are indebted to Colin Landon, Gregg Radtke and Austin Minnich for many useful comments and discussions. 
This work was supported in part by the Singapore-MIT Alliance. J-P. M. P\'{e}raud gratefully acknowledges financial support from 
Ecole Nationale des Ponts et Chauss\'{e}es and 
the MIT Department of Materials Science and Engineering through a Graduate Fellowship.

\newpage

\appendix

\section{Numerical data for scattering rates}

In the simulations presented here we use data for the dispersion relations and for the relaxation times of phonons in Al and Si. Dispersion relations are adapted from the experimentally measured dispersion in the [100] direction (\cite{Stedman66} for Al, \cite{link_Si,minnich} for Si).

For Al, as in \cite{Minnich11_not_submitted,minnich}, we assume a constant relaxation time chosen to match the desired lattice thermal conductivity. We therefore take the value
\begin{equation}
\tau_{Al}=10^{-11} \mathrm{s}
\end{equation}

For Si, we use the expressions from \cite{henry08}, with constants from \cite{minnich}. Relaxation times for acoustic modes are given by

\begin{center}
\begin{tabular}{l l}
\hline
phonon-phonon scattering, LA  &$\tau^{-1}_{L}=A_L \omega^2 T^{1.49} \exp \left(\frac{-\theta}{T} \right)$ \\
phonon-phonon scattering, TA  &$\tau^{-1}_{T}=A_T \omega^2 T^{1.65} \exp \left(\frac{-\theta}{T} \right)$ \\
impurity scattering  &$\tau^{-1}_{I}=A_I \omega^4$  \\
boundary scattering  &$\tau^{-1}_{B}=w_b$ \\
\hline
\end{tabular}
\end{center}
\vspace{1cm}
where the constants take the following values
\vspace{.5cm}
\begin{center}
\begin{tabular}{c c c c c c}
\hline
Parameter &$A_L$ & $A_T$ & $\theta $ & $A_I$ & $w_b $ \\
Value (in SI units) & $2 \times 10^{-19}$ & $1.2 \times 10^{-19}$ & $80$ & $3 \times 10^{-45}$ & $1.2 \times 10^6$\\
\hline
\end{tabular}
\end{center}
\vspace{.5cm}
The total relaxation time for a given polarization is obtained using the Matthiessen rule 
\begin{equation}
\tau^{-1}=\sum_{i} \tau^{-1}_i
\end{equation}
\vspace{1cm}

Optical phonons in Si are considered immobile (Einstein model). 
Einstein's model states that the contribution of optical phonons to the vibrational energy per unit volume 
in a crystal is given by \cite{chenbook}
\begin{equation}
U=\frac{N_p N' \hbar \omega_E}{V[\exp(\hbar \omega_E/k_b T)-1]}
\end{equation}
where $N_p=3$ is the number of polarizations, $N'=1$ is the number of optical states per lattice point, $\omega_E$ is the Einstein radial frequency ($\omega_E=9.1\times 10^{13} s^{-1}$ \cite{link_Si,minnich}), V is the volume of a lattice point (with a lattice constant $a=5.43\AA$, $V=a^3/4=4\times 10^{-29} m^3$).

For the relaxation time of optical phonons, we use the value \cite{Klemens66}
\begin{equation}
\tau_O =3 \times 10^{-12} \mathrm{s} 
\end{equation}

\section{Derivation of the transient ballistic 1D solution}
\label{sec:ballistic}

\begin{figure}[htbp]
\begin{center}
\includegraphics[width=.8\textwidth]{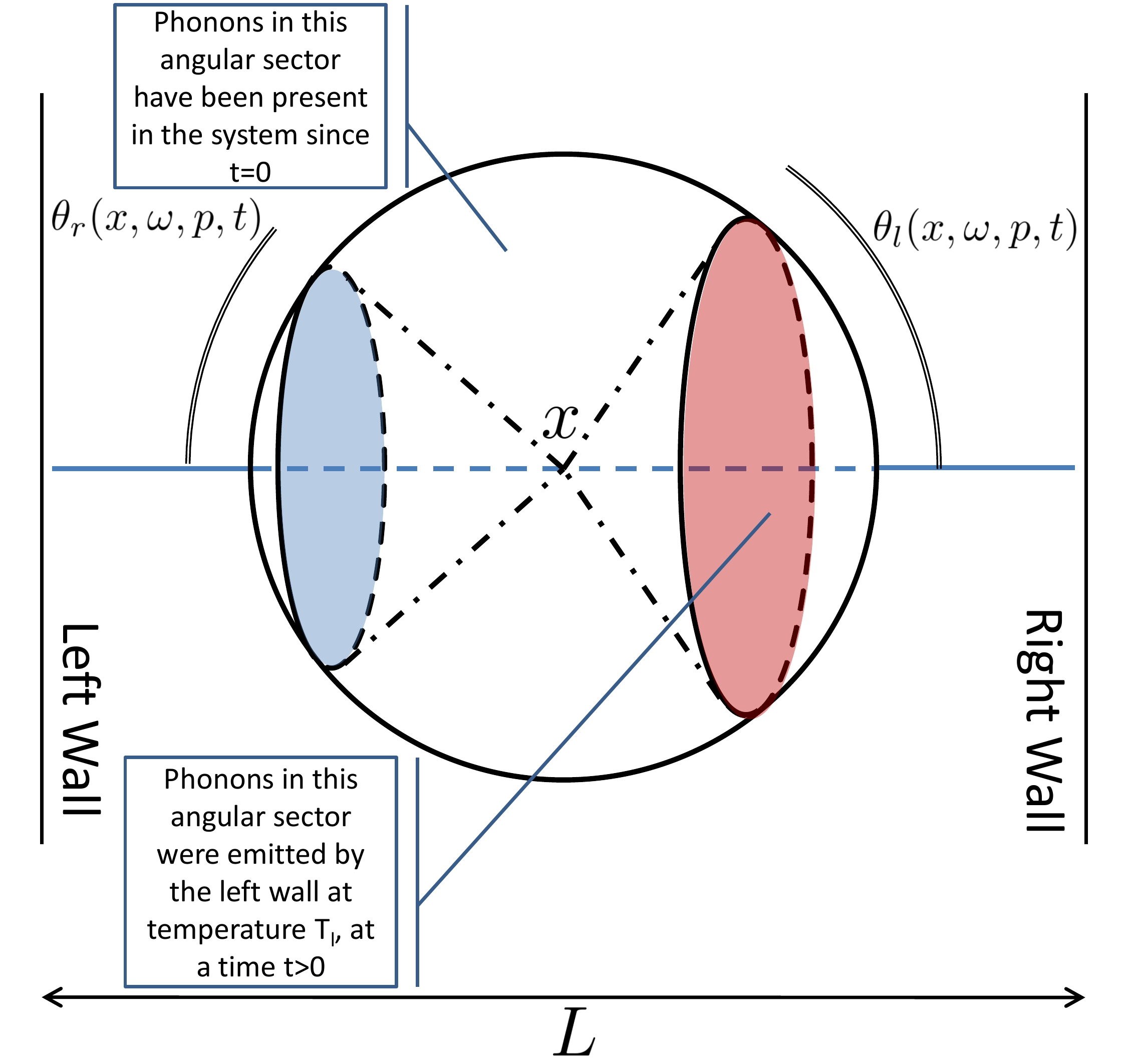}
\caption{At a given point in space, the solid angle can be divided into three distinct regions in which the distribution 
of phonons is known; here, $\theta_{l}$ is given by $\cos(\theta_{l})= x/ (V_{g}(\omega,p) t )$, while $\theta_{r}$ is given by
 $\cos(\theta_{r})= (L-x)/ (V_{g}(\omega,p) t )$}
\label{fig:ballistic_angular}
\end{center}
\end{figure}

Following the impulsive change of temperature at the walls from $T_0$ to $T_l=T_0+\Delta T$ and 
$T_r=T_0-\Delta T$, thermalized phonons at temperature $T_r$ and $T_l$ 
are emitted from the ``right'' and ``left'' wall, respectively (see Figure \ref{fig:ballistic_angular}). 

For some arbitrary location $x$, for a given frequency, polarization and time, the angular space can be divided into 3 
distinct domains characterized by two angles $\theta_r(x,\omega,p,t)$ and $\theta_l(x,\omega,p,t)$ as depicted in 
Figure \ref{fig:ballistic_angular}. Phonons described by $0<\theta<\theta_l$  were emitted by the
left wall at a time $t>0$. Phonons described by $\theta_l<\theta<\pi-\theta_r$ have been
present in the system since t=0. Phonons described by $\pi-\theta_r<\theta<\pi$  were emitted by the
right wall at a time $t>0$. 

The energy can therefore be written as
\begin{multline}
E_V (x,t) = \frac{1}{2} \sum_{p} \left\{ \int_{\omega}\int_{\theta = 0}^{\theta_{l}(x,\omega,p,t)}e^{eq}_{T_{l}}(\omega) D(\omega,p) \sin(\theta)d\theta d\omega ... \right. \\ + \int_{\omega}\int_{\theta = \theta_{l}(x,\omega,p,t)}^{\pi-\theta_{r}(x,\omega,p,t)}e^{eq}_{T_{0}}(\omega) D(\omega,p) \sin(\theta)d\theta d\omega ...\\ \left. + \int_{\omega}\int_{\theta = \pi-\theta_{r}(x,\omega,p,t)}^{\pi}e^{eq}_{T_{r}}(\omega) D(\omega,p) \sin(\theta)d\theta d\omega \right\}
\label{eq:ballistic_formula}
\end{multline}

From geometrical considerations
\begin{align}
\cos\left(\theta_r(x,\omega,p,t) \right)&=\text{min}\left(1,\frac{L-x}{V_{g}(\omega,p)t} \right) &=1-\left(1-\frac{L-x}{V_g(\omega,p)t}\right) H\left(1-\frac{L-x}{V_g(\omega,p)t}\right) \\
\cos\left(\theta_l(x,\omega,p,t) \right)&=\text{min}\left(1,\frac{x}{V_{g}(\omega,p)t} \right) &=1-\left(1-\frac{x}{V_g(\omega,p)t}\right) H\left(1-\frac{x}{V_g(\omega,p)t}\right) 
\end{align}
where $H$ is the Heaviside function.
Proceeding to the integration in $\theta$, the energy density is given by
\begin{multline}
E_V(x,t) = \frac{1}{2} \sum_{p} \left\{ \int_{\omega}\left(1-\frac{x}{V_g(\omega,p)t}\right)H\left(1-\frac{x}{V_g(\omega,p)t}\right)e^{eq}_{T_{l}}(\omega) D(\omega,p)d\omega \right. \\ +\int_{\omega}\left(1-\frac{L-x}{V_g(\omega,p)t}\right)H\left(1-\frac{L-x}{V_g(\omega,p)t}\right)e^{eq}_{T_{r}}(\omega) D(\omega,p)d\omega \\+ \int_{\omega}\left[1-\left(1-\frac{x}{V_g(\omega,p)t} \right)H\left(1-\frac{x}{V_g(\omega,p)t} \right)\right]e^{eq}_{T_{0}}(\omega) D(\omega,p)d\omega \\ \left. + \int_{\omega}\left[1-\left(1-\frac{L-x}{V_g(\omega,p)t}\right)H\left(1-\frac{L-x}{V_g(\omega,p)t}\right)\right]e^{eq}_{T_{0}}(\omega) D(\omega,p)d\omega \right\}
\label{eq:ballistic_breakdown}
\end{multline}
The temperature $T=T(x,t)$ is obtained by numerically finding the Bose-Einstein distribution corresponding to this energy density.

Using the Debye model and considering small temperature changes ($|T_r-T_0|<<T_0$ and $|T_l-T_0|<<T_0$), 
the resulting temperature field can be expressed in a simpler form. The first assumption allows the removal of the frequency and polarization dependence on the group velocity, 
while the second assumption allows the linearization of the Bose-Einstein terms in the integrals. Several simplifications can then be carried out to 
yield the following expression for the temperature field
\begin{equation}
\Delta T(x,t)=\frac{1}{2}\left(1-\frac{x}{V_{g}t} \right)H\left(1-\frac{x}{V_{g}t} \right) \Delta T_{l}+\frac{1}{2}\left(1-\frac{L-x}{V_{g}t}\right)H\left(1-\frac{L-x}{V_{g}t}\right) \Delta T_{r}
\label{eq:ballistic_closed_form}
\end{equation}

\bibliographystyle{ieeetr}
\bibliography{phonon_bib}
\nocite{*}

\end{document}